# Conceptualization and experimentation of asset market with price manipulation

**Pau Fonseca i Casas[1], Aarón Montero Montero [2]**
[1]Corresponding Author. Universitat Politécnica de Catalunya. Statistics and Operations Research Department. Campus Nord. Jordi Girona 1–3 08034, Barcelona, Spain {pau.fonseca@upc.edu}. Tel. + 34 (93) 4017732; Fax. (+ 34) 93 4015855.
[2]Universitat Oberta de Catalunya, Barcelona, Spain {amonteromon@uoc.edu}.

## 1. Abstract

The primary goal of this work is to reproduce the behavior of a human trader, detailing his or her psychological processes to understand the effects of his or her decisions on the final value of an asset. The second goal is to use a formal language to detail this behavior as a tool for improving and simplifying the communication between all actors involved in the project, such as specialists from disciplines as diverse as computing, economics and psychology. As a starting point, we use a paper that shows an experiment that analyzes the influence on other traders' behavior when an agent handler and a trading robot attempt to distort the market. This work reproduces this experiment, using virtual traders that belong to a multi-agent simulation model, showing the feasibility to reproduce complex human behaviors and showing the convenience of use formal and graphical languages to simplify the understanding and the validation of the complex behaviors involved in an economic process.

## 2. Introduction

The analysis of the economical behavior using computational techniques is a well-known area where several techniques are applied to try to predict the system behavior (Wang, Quek, & Ng, 2016). The integration of psychological factors in economic decision-making has been studied by several authors, such as (Brandstätter, 1993), (Lo, Repin, & Steenbarge, 2005), (Rabin, 2002), (Mehlenbacher, 2009) or (Langley, Bradshaw, Simon, & Zytkow, 1987). More specifically, the simulation of economic behavior has been studied to emulate the behavior of the markets. When we want to analyze the relationship between some rules applied to some entities and the emergent behavior, Multi-Agent-Based Simulation (MABS) (Davidsson, 2001) becomes an effective method (Russell & Norvig, 2011), (Wan, Hunter, & Dunne, 2002). Financial markets, and in our case the behavior of financial markets due to the behavior of the traders, belongs to this kind of model in which the relationships between entities cause an emergent behavior that often generates wealth. A very detailed model that is based on the behavior of the agents can be reviewed (LeBaron, Arthur, & Palmer, 1999). This model uses MABS to reproduce certain time-series features of the real markets. In (Takahashi & Terano, 2003), a MABS is presented which is used to analyze how asset prices are affected by investor behavior. The behavior of the investors is rational (based on traditional financial theories) and/or irrational (based on Behavioral Finance). Hoffmann (Hoffmann, Jager, & Von Eije, 2007) presents a MABS model in which the agents are based on real theories to emulate real stock market behavior. More recently a work suggests that is not needed to implement learning procedures on trader's agents to represent their behavior (Rayner, Phelps, & Constantinou, 2014).
In our work, we use real market and real psychological theories to represent the agent's (our trader's) behavior. From this representation, we try to validate the model by defining an experiment equivalent to the experiment developed by Veiga and Vorsatz (Veiga & Vorsatz, 2008) and, thus, compare whether the obtained results from our experiment are similar to those obtained in the referred paper. In our approach, we follow the ideas of (Langer, et al., 2021) defining graphically the model assumptions. This simplifies the validation process done by the different experts that must assure the correctness of the model, the developers on the validation process.

The psychosocial components of the behavior of any operator in the stock market play a very important role when we try to understand the evolution of the financial markets. It is necessary to understand that a decision to purchase and sell shares or securities is a behavior-based decision made with certain expectations: maximizing profit. What happens in the mind of the trader when deciding to buy or not buy, to sell or not sell securities? Cognitive psychology was developed during the second half of the 20th century as a result of behavioral psychology's limitations in explaining cognitive processes (which are considered the key to understanding human behavior). This psychological approach coincides with the beginning of computer science, and certain parallels have been established between the brain and the computer. However, while in both cases information is processed (Ballesteros S. , Psicología General: Un enfoque cognitivo, 2000), the computer metaphor initially considered that mental processes ran sequentially, as when we execute algorithms. Subsequently, the metaphor evolved into connectionism, based on the neuronal structure of the brain, an approach that allows parallel processing (Pitarque & Algaradel, 1991).
In the analysis and perception of a specific situation, *external data* are received. There also exists some “internal data” that belongs to the subject's brain. Internal data is important because it is used to filter the detection of *external data* and conditions its understanding. We do not see and hear the things that *really* exist, but we understand the things that are coherent with our way of seeing and hearing the world, *our world*. In that way, each trader elaborates on their own stock-market situation. At the same time, they try to imagine the mental elaborations of the other traders that compete with them. The analogy in computer science is often based on the use of Multi-Agent Simulations (MABS).
But what determines the different decision-making behavior of traders? Intelligence is the effectiveness of the human mind in solving problems of understanding the world and our relationship with it. Some psychologists like Howard Gardner talk about multiple intelligence (Gardner, 2006), distinguishing eight types or aspects of intelligence: linguistic, logical-mathematical, spatial, musical, bodily-kinetic, intrapersonal, interpersonal, and naturalistic. To solve a given problem, all these aspects are involved; but some facets of intelligence are more influential than others, depending on the problem or the individual. For the problems that a trader needs to solve, the primary facets are: logical-mathematical, interpersonal, intrapersonal, and naturalistic. Throughout the problem-solving or decision-making process, different processes are involved. One process is the conscious and rational analysis of the variables considered to be causative or intervening factors. But it is obvious that before we reach solutions obtained by this analytical process, there are other *intuitive* solutions or hunches (Kahneman & Tversky, 1972).
Within this description of personal behavior, some are inclined to consider the personality as "stable", which would enable predictions of their behavior. This would be the current psychology of personality traits. According to the score achieved in different traits, the subjects will be assigned to one personality type or another. But others are more inclined to view the evolution of the personality, understanding the personality as "dynamic." Psychologists interested in responding to problematic situations in clinical practice, such as Carl Rogers (Rogers & Kramer, 1995), use client-centered therapy and
the development of personality.
Most personality tests are based on self-reporting questionnaires or inventories where individuals compile their behaviors, feelings, and opinions. The suspicion that the subject uses the questionnaire as a tool for forming a socially correct view of his or herself decreases its validity. Moreover, the psychological state of the subject who performs a personality test may differ from one moment to another, which also compromises its reliability. One of the theories applied in the tests is that of personality traits. We define a feature as "a tendency or style of behavior" that has some stability across time and situations, selecting those features that best define the setting of a person and the society in which he or she is immersed. Gordon W. Allport (Allport, 1963) understands behavioral traits, such as the possible margins, which can be activated as required by the situation. This defines personality as a structure that reflects the individually different traits of an organization.
This project has two main objectives: first, to reproduce the behavior of a human trader[1], detailing his or her psychological processes and understanding the effects of his or her decisions on the final value of an asset (based on (Veiga & Vorsatz, 2008) work); and, second, to use a formal language to define this behavior. To validate our assumptions, we implement a model that follows the structure of the experiment described in (Veiga & Vorsatz, 2008). Different elements are taken into consideration, like the existence of a manipulative agent that tries to modify the asset value. The manipulative agent is represented by an intelligent agent that tries to obtain profit from this manipulation.
This paper is organized as follows. In section 3, the trader’s personality is detailed from the point of view of psychology to define the main traits to be used in our model. In section 4, we detail the assumptions to be used in our model according to the previous section. In section 5, we detail the formalization and the implementation of the model, while in section 5.2 we discuss the main concerns related to the Validation and Verification of this model. Section 6 shows the results and section 7 compares the results with those obtained (Veiga & Vorsatz, 2008). Finally, section 8 details some concluding remarks.

## 3. Defining the trader’s personality

In our model, we try to outline the key aspects of the trader’s personality, which define his or her behavior in the experiment conducted (Veiga & Vorsatz, 2008). Among the different tests that can be used to describe the personality of people not affected by psychological pathologies, we can use two tests based on the factorial analysis of the traits: the 16PF[2] of Cattell (Cattell, The Scree Test for the Number of Factors, 1966) (Cattell, Eber,

[1] A trader is person or entity in finance who acts in the capacity of agent, hedger, arbitrageur, or speculator in buying and selling financial instruments such as stocks, bonds, commodities and derivatives..
[2] 16PF: Sixteen Personality Factors

& Tatsuoka, Handbook for the Sixteen Personality Factor Questionnaire (16 PF), 1980) and the EPQ[3] of Eysenck (Eysenck, 1991). To describe our trader's personality, 16PF has 185 items. The application time for 16PF is about 45 minutes. These 16 primary factors can be combined into five global scales: Extraversion, Anxiety, Tough-mindedness, Independence, and Self-control.
In line with this type of survey is the EPQ, which assesses three dimensions:

1. Extraversion - Introversion
2. Neuroticism - Stability
3. Psychoticism – Sensitivity

EPQ needs between 15 and 30 minutes to be applied.
The trait of extraversion-introversion describes its openness to social contact. Extroverts are expansive, uninhibited, easy to contact socially, and impulsive, and they prefer movement and action. On the other hand, introverts are thoughtful, orderly, and reserved in terms of social contact.
The trait of neuroticism-stability (anxiety-adjustment) describes emotional balance. Neurotics are emotionally hypersensitive and frequently suffer diverse somatic disturbances (headaches, digestive disorders, insomnia, backaches), which are reflections of their emotional tension or anxiety.
The trait psychoticism-sensitivity (hard-sensitivity) describes the quality of feeling concerning human relations. Psychopaths are cold, cruel, unpleasant, and inhuman; they do not empathize with the feelings of others, making them insensitive beings. They take advantage of others and feel no remorse for their cruelty (Ballesteros R. F., 1984).
Let us look at some aspects of personality that can affect the behavior of a trader, to finally decide what can be considered applicable to our project.

### 3.1. The trader and his or her resistance to stress

Occupational stress occurs when the individual experiences constant emotional stress due to a perception of risk or danger that forces them to be on alert. It is not the workload but the lack of expectation of carrying it out safely and successfully. In works where already known procedures can be applied successfully, individuals do not experience anxiety and therefore no stress. In these situations, the mental energy expenditure is minimal. In contrast, novel situations are more mentally demanding. If novelty is added to the risk of failure, mental effort, and emotional stress are higher.
In the case of traders, we have a work situation that requires innovative responses, as the market situation is changing from one moment to another. The traders have to take into account the evolution of the values regarding that which they wish to buy or sell. Forecasts can work in the short or long term, according to the instructions received from their constituents. In the past, knowledge of financial market evolution was a source of information for forecasting but was never repeated the same way. In changing situations, tendencies can be distinguished in moments of stability and in times of crisis, but where there are changes there are also unpredictable trends. In times of market crisis, the stress level is higher because there is less reliance on the projections. The trader's stress is a function of their risk. Their risk can be measured by their economic losses, losses in self-esteem and professional standing, or their losses in job security. But everyone sees risk situations differently. Experienced traders suffer less stress than novices. The traders with an outgoing personality experience less stress than introverts, who are more calm and cool than neurotics. People prone to anxiety avoid situations of uncertainty, such as the trader profession.
The relationship between stress and job performance is contained in the Yerkes-Dodson Law (Nelson, 2007). Introverted subjects are more likely to experience anxiety and reach the critical point earlier than extroverted subjects, which means they have the most impaired performance in situations involving sustained stress. In short, introverts withstand less pressure. For example, an introverted, conservative trader will take fewer risks in any situation, but especially in crises. Instead, extroverts are more likely to take risks. The decline in the performance of a trader can be seen in the reduction of data used, an increase in the time needed to make a decision, and an increase in the use of heuristics.

### 3.2. The trader and his or her moral development

All professional activities are performed within norms, but in addition to established social norms, personal values exist. One of the most influential moral development theories has been developed by the Swiss developmental psychologist, Jean Piaget, whose theory has been updated and refined by his student Lawrence Kohlberg (Palomo González, 1989). Kohlberg divides this development into three levels, each of which is divided into two stages:

- Level 1: Pre-conventional: Stage 1: Acceptance of the rules by fear of punishment. Stage 2: Acceptance of the rules for mutual benefit.
- Level 2: Conventional: Stage 3: Conduct guided by expectations about relevant people we have around us. Stage 4: Acceptance of the rules set by society as a common good.

[3] Eysenck Personality Questionnaire

- Level 3: Post-conventional: Stage 5: The behavior is based only on the rules and the recognition of the rights granted to all beings. Stage 6: Behavior is guided by universal ethical principles and recognition of the dignity of others.

Instruments used to assess the level of moral development is the resolution of "moral dilemmas". Each person takes his route and is situated at a certain level.
The moral development of the trader determines how they will use the financial market. If they are only guided by their benefit (level 1) they do not mind buying shares of companies that do not respect human rights or the environment, and they may even take actions that result in serious damage to others. The lack of scruples in the financial markets, in terms of the benefits of selfishness theory, was one of the factors that contributed to the 2008 financial crisis, coupled with weak regulatory and control mechanisms.

### 3.3. The trader and his or her analytical and intuitive competence

In the same way that there are cognitive styles in the process of learning, there are cognitive styles in decision-making. Individuals who process information predominantly in a reflective, analytical and rational way seek solutions that are reasonably explained. On the other hand, individuals who are guided more by intuition or by creative solutions are not able to provide rational explanations for their decisions, but instead, they are confident in the strength of their intuition. These intuitive decisions can be based on automatic data processing based on previous experiences. It is sometimes described as having a critical eye, in the case of some doctors or certain political leaders during times of crisis. The most commonly used strategies are: a) trial and error, b) hypothesis testing, c) algorithms and d) heuristics (Ballesteros & García, Procesos psicológicos básicos, 2005).
The trader profession seems to be reserved for thoughtful and analytical individuals; but we cannot exclude traders who are also of an intuitive type, who are heavily influenced by their confidence in their intuitive potential.

### 3.4. Prejudices and stereotypes of the trader

In the process of analyzing the variables involved in the power of a certain financial value, a trader's prejudices and stereotypes interfere with rational analysis. For example, in the European debt crisis, a consequence of the Great Recession of 2008, financial markets dealt with objective data as well as prejudices and stereotypes that the Anglo-Saxon culture holds toward Mediterranean culture. Just as there are prejudices and stereotypes about race and culture, there are stereotypes and prejudices about the economy of a company or country. These stereotypes sometimes are justified in the interest of profit (Morales, Moya, Gaviria, & Cuadrado, 2007) (Tajfel, 1984).
The traders who have prejudices and stereotypes will make decisions to buy or sell according to these beliefs and stereotypes. This will lead to different treatments for financial assets, taking into account the pertinent countries or companies.

### 3.5. The resistance to social pressure

Related to prejudice and stereotypes is the submission to social pressure. Stereotypes are a kind of social pressure through social beliefs. But social pressure has a more immediate aspect connected with the environment surrounding the individual. For a trader, the surrounding environment formed by the other traders may create a certain amount of pressure, depending on their resistance to social pressure. Individuals who are guided by specific criteria such as the opinions and behavior of others experience less anxiety and stress when their behavior disagrees with those around them. Consequently, individuals who need social approval are more likely to modify their behavior according to social pressure.
Resistance to social pressure is positively correlated with the perception of efficacy. Individuals who are perceived as capable and efficient place the Locus of Control (LOC[4]) on themselves. The psychological construct called Locus of Control refers to the location of the main cause of the achievements or reinforcements. In some individuals, LOC is located in external agents or the situation. Other individuals locate the LOC in their ability and effort. Subjects who lose every expectation that their behavior might influence their failure or success in achieving something may eventually become inactive as a result of so-called "learned helplessness" (Peterson, Maier, & Seligman, 1995). To measure this trait, Rotter (Rotter, 1990) developed the Scale of Internal-External Control. This concept of LOC relates to the causal attribution processes of individuals. Causal attribution studies were initiated by Heider and Simmel in 1944. Heider's theory identifies four types of cases: two personal (ability and motivation) and two situational (Locus of Causality [internal – external]) and Stability [stable – unstable]) (Moreno Bermúdez, 2000). This view is particularly important in times of crisis because social pressure resulting from the behavior of others is important in determining the power of market values.

### 3.6. The trader and risk-taking

Resistance to stress and social pressure is related to perceived self-efficacy, in that it allows us to classify people in terms of their tendency to take risks. Risk-taking in financial markets is also related to the cost of bad debts.

[4]LOC may be regarded as a personality trait and has been studied by Rotter, Chance, and Phares (1966, 1972).

One of the criticisms of the financial system in the 2008 crisis was the low personal risk assumed in operations that involved high financial risk.
In a financial system where the effectiveness of stock trading directly affects the trader, his or her professional behavior is in part driven by his or her ability to take risky decisions. The assumption of risk is higher in extroverted, psychopathic, and independent individuals than in introverted, sensitive, anxious, and dependent individuals. There are other ways to take risks, which are linked to a lack of knowledge about a situation or thoughtless behavior.

## 4. Modeling assumptions

In this section, we detail the assumptions used in our model. They are divided mainly into two sections: the first is related to the traders while the second is related to the experimental framework used.

### 4.1. Trader modeling assumptions

Although all the features noted above influence the way a trader does his or her work, to simplify our study we will choose one feature that reflects his or her practice (expert - novice) and only three personality traits:

1. Emotional Adjustment - Resistance to stress (calm - anxious)
2. Social dependency (independent - dependent)
3. Cognitive style (analytical - intuitive)

The main problem lies in finding some operators who reflect these traits in the different processes carried out by the trader in their work. As this project aims to study the influence of psychological types not in the actual market but a computer simulation, we dispense with the experimental verification of how these personality traits influence the behavior of real traders. Operators will be:

1. The **delay (time)** between trading
2. The **scope of the analysis** of data before deciding.
3. The **extent of risk** assumed when formulating their offers for sale and purchase requests.

#### 4.1.1. First trader modeling assumptions 1: the delay time

To model how the brokers, behave over time, one can use an activity scanning approach in which time is divided into several time steps and, in each one of these time steps, we analyze what to do regarding the model processes (Withanawasama, Whighama, & Crack, 2013). We are following a discrete simulation approach, which needs details about the next event (success) to be executed. Since this implies a causality order in our model, the delay time is an aspect that must be defined. It is not relevant for us to provide our model with exact details at this time, only to depict that –compared to others– the **delay time** will be greater for the traders who are rookies, anxious, dependent, and analytical. We establish three levels of delay: minimum, medium, and high. The times are based on the assumptions described in (Veiga & Vorsatz, 2008). The subjects trade the asset for five minutes. It is not our intention to represent the delay time in detail, but to depict three proposed response times that lead to performing more actions if less delay is applied:

1. Minimum delay: 3 seconds.
2. Average delay: 8 seconds.
3. High delay: 16 seconds.

The choice of delay time for a given operation will be obtained depending on that level. Given the performance curve at work, the delay will also be modified by the level of fatigue. We introduce a correction factor for these values throughout the various rounds. We apply the Yerkes-Dodson Law (Yerkes & Dodson, 1908), where two rounds are needed to gain a high delay. For the traders who possess the trait of high anxiety, that performance will begin its descent from round seven. In the first round, the values will increase by 2 for the minimum delay, by 3 for the average delay, and by 7 for the high delay. In the second round, the values will increase by 1 for the minimum delay, by 1 for the average delay, and by 3 for the high delay. The seventh-round only applies to the traders who have high anxiety status. It is applied as follows: the values will increase by 1 for the minimum delay, by 1 for the average delay, and by 3 for the high delay. The eighth round, again, only applies to traders with high anxiety, increasing the values by 2 for the minimum delay, by 3 for the average delay, and by 7 for the high delay. Fig.1 and Fig.2 present the delay time for the low anxiety profile and the high anxiety profile, respectively.

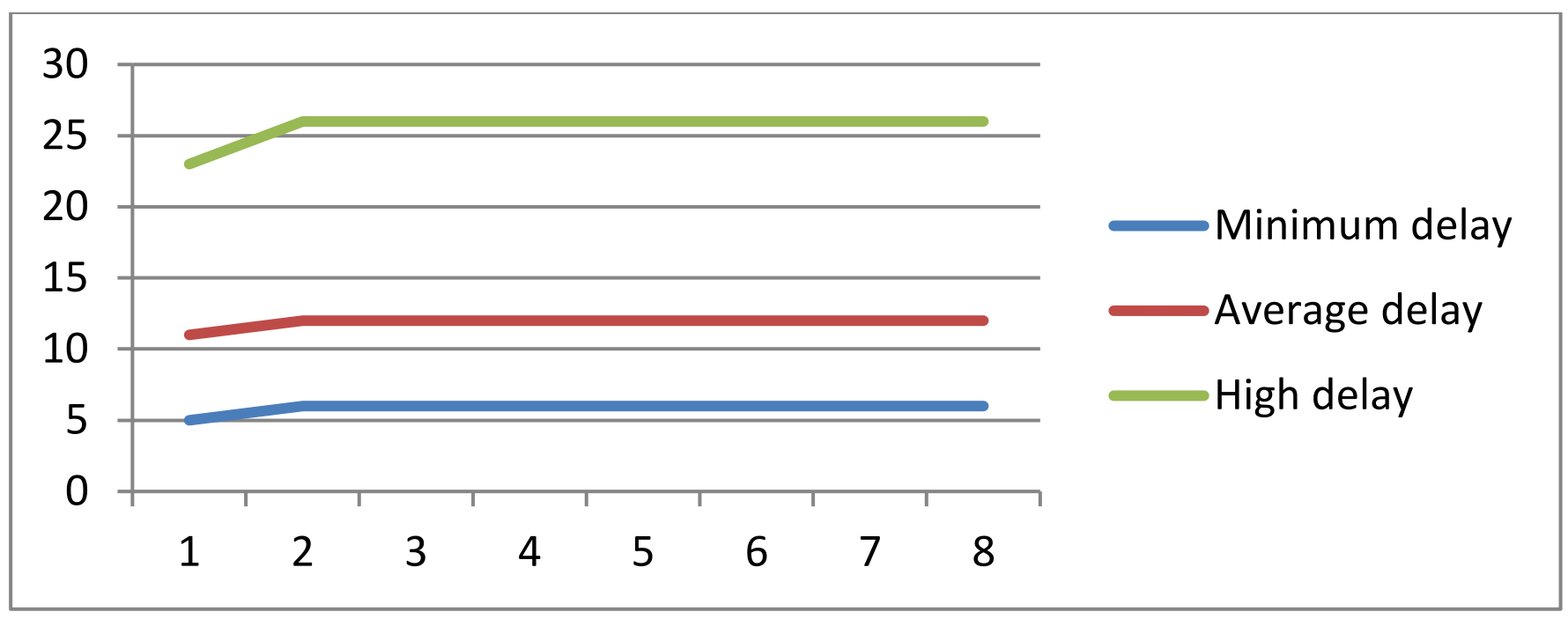

Fig.1. Delay time for low anxiety profile. On the ordinate axis, the delay time is shown. In the abscise axis, the number of the round.

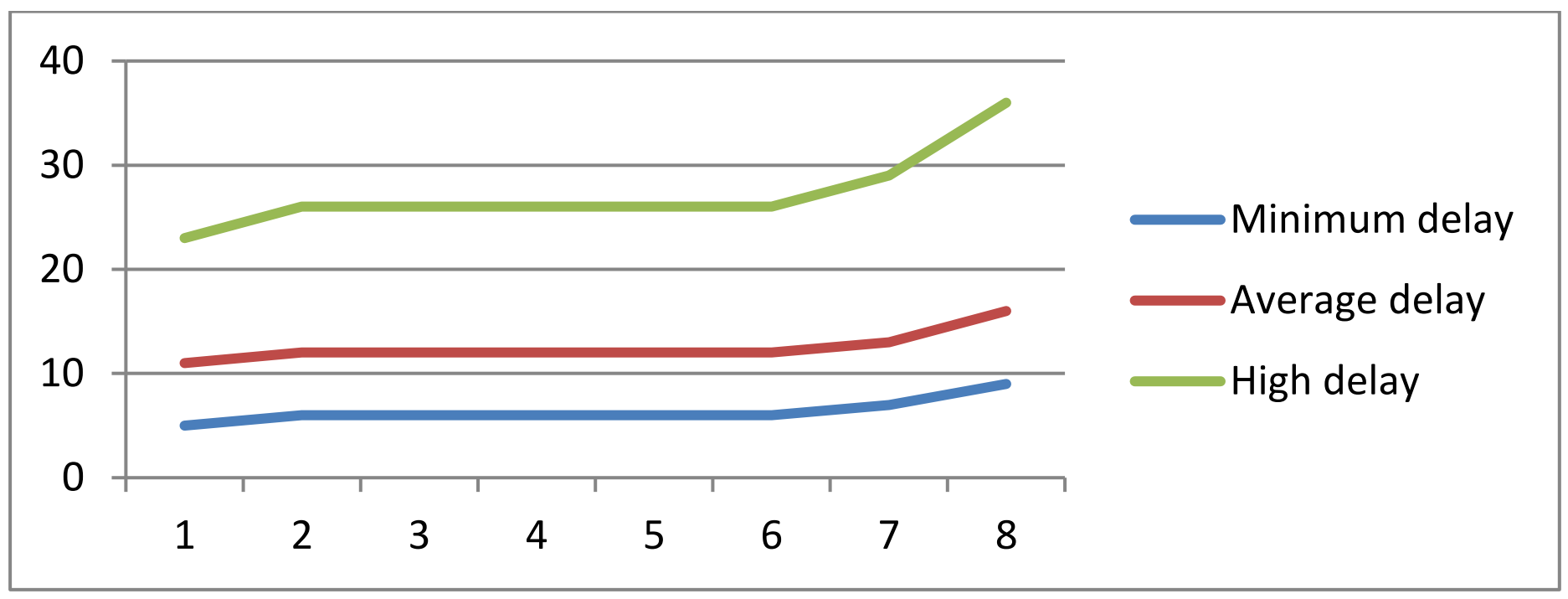

Fig.2. Delay time for high anxiety profile. On the ordinate axis, the delay time is shown. In the abscise axis, the number of the round.

#### 4.1.2. Second trader modeling assumptions: the scope of the analysis

The **scope of the analysis** (data to consult before making the decision) will be higher for expert, calm, independent, and analytical traders. The data will be provided on request according to firm, market, and history. Agents conduct a stock market technical analysis if the trader has the proper experience for doing so. We use the techniques described in (Cárpatos, 2006). This technical analysis consists of the following information:

1. **Dow Theory** (Dow, 1844). The Dow Theory analyzes the support and resistance lines, which mark a turnaround in the case of being overwhelmed. This theory did not apply to our system, since in our case the traders know the maximum and minimum value of an action. Thus, the behavior would be the opposite: it should be near a support line of resistance (in our case, 100 ECU and 220 ECU, respectively; hence the trend continues and the price could be stabilized at that value). Nevertheless, since our system analyzes a special case, the Dow Theory and our particular experiment will coexist naturally. The choice of one or the other should be determined by stipulating that the company can range only between two constants: maximum and minimum values.
2. **Relationship between market value and volume of shares**. The changes in market value and volume of shares determine if the changes go in the same direction (i.e., increase or decrease). If there is a divergence, we face a change in trend. If the stock goes down in volume but market value increases, it could represent an indicator of a future bearish trend; otherwise, it could represent the next uptrend. In our case, as a restraint of the experiment, it is not possible to perform operations on more than one action at a time. This relationship was not taken into account when predicting the market, although it could be considered useful in future implementations.
3. **Elliot waves** (Precher & Frost, 1989) combined with the theory of contrary opinion. In Elliot waves, we determine the wave where we are at all times. According to this, and knowing what their trend is and that this will change in the next wave, we assume that this change is imminent, as stated by the theory of contrary opinion. So, if the wave in which we find ourselves indicates that the value of the stock is down, we assume that this situation will change soon and therefore buy; whereas, if the stock value is rising, we assume that it will soon decrease, in which case we sell.
4. **Trend**. Following the maximum and minimum price values of the company, we will determine the trend, whether it be bullish, bearish, or stable.
5.

With this information, the agent will be able to predict the future market. In the case that the trader does not have the experience to develop this analysis, the trader will base their buying and selling orders only on whether the stock value goes up or down at the moment. This scheme is also used by the expert trader when market signals are not entirely clear; but the opposite case can be made for the expert trader, who uses the theory of contrary opinion. That is, while the inexperienced trader buys when the shares rise and sells when stocks go down, the experienced trader will assume that the inexperienced expert is wrong, then sell when stocks rise and buy when they fall.
Once a prediction about the market is made, and if the trader sees the possibility to sell or buy, he or she will launch an offer to buy or sell whenever conditions permit, i.e., if the trader owns enough credit or tangible actions to make the offer. After determining the price for which the agent is willing to buy or sell a stock, it remains only to analyze the information that reaches the market run, i.e., bids that have been released by other traders. In this way, we can bring back our bid to get more profit or more easily find a buyer or seller. Finally, once an offer is improved, it is necessary to ensure that the trader is still able to respond with a credit or actions if the negotiation is closed.
Regarding the other agents, only the market agent will change his or her operation. He or she will determine the best and second-best buyers and sellers in the market, so that traders can obtain information from the ring of traders to improve their offers and the closure of their operations. For a description of the agents that compose the model, Fig.3 can be reviewed (in section 5 we detail this).

#### 4.1.3. Third trader model assumptions: the extent of risk

The **extent of the risk** will be greater in rookies, independent and intuitive traders. Three levels of risk also apply, relating to 1) the timing of buying and selling, 2) the ask price as fixed to the bid price, and 3) stock conservation.
Low-risk profiles:

- Buy if the value is below the intermediate value and tends to rise.
- Set the "bid price" to rise between 1 and 2% of the lowest sell offer.
- Sell if the last value remains at least 10% higher than when the trader bought, or the value is above the intermediate value and begins to drop.
- Set trader ask price to undercut between 1 and 2% of the higher purchase request.
- Limit the use of the loan to 50%

Medium risk profiles:

- Buy if the value is below the intermediate value and tends to rise.
- Set the "bid price" to rise between 3 and 6%, the lowest sell offer.
- Sell if the last value retained is at least 7% higher than when the trader bought, or the value is above the intermediate value and begins to drop.
- Set trader ask price to undercut between 3 and 4% of the higher purchase request.
- Limit the use of the loan to 75%

High-risk profiles:

- The trader will buy if the value is going up, or if it has gone down close to its minimum contribution
- Set the "bid price" to rise between 6 and 10% of the lowest sell offer.
- Sell if the last value remains at least 5% higher than when the trader bought, or the value is above the intermediate value and begins to drop.
- Set trader ask price to undercut between 5 and 8% of the higher purchase request.
- Do not limit the use of the loan

### 4.2. The experimental framework assumptions

Once we have the theoretical definition for the trader's traits, the modeling assumptions, and the elements needed to execute the model, we go further to define the experimental framework that allows us to execute the model and perform the validations. The category of professional experience will be divided into two groups: experts (seniors) and novices (juniors). We estimate that the number of expert traders working in the financial market is clearly above that of the novice, as in most professions. Traders with 2-3 years' experience will be considered as junior traders and those with 5 years or more as senior traders; hence we set 80% seniors and 20% juniors.
The category of emotional adjustment (calm - anxious) is distributed according to the following criteria: serene 50%, medium anxiety 30%, and high anxiety 20%. The reason for giving more weight to serene is based on the logical assumption that people with high anxiety tend to avoid stressful situations; so, with the understanding that the financial market generates highly stressful situations, we assume that anxious people tend to avoid this kind of job. The category of social dependency is divided according to the following criteria: 25% independents, 50% with normal dependence, and 25% with high dependence. For the analytical style category, we apply similar criteria to the anxious category. For that, we assign 50% to analytical subjects, 30% to intermediate intuitive subjects, and 20% to highly intuitive subjects.
At this point, we have finished developing the theoretical behavior of a broker following the conditions imposed in the article by Veiga and Vorsatz (Veiga & Vorsatz, 2008). However, we are not interested in the theoretical simulation of a broker, but human behavior, so that the simulation results will be as close as possible to what happens in reality. We recognize that human behavior is extremely complex, even when taking into account the many factors and variables that can be found in an area as complex as the financial markets. Therefore, we try to extract a portion of this behavior from a simple model with only the most important personal characteristics, so that we can obtain an approximate overall performance rather than get mired in complex and inaccurate behavior. Therefore, we want to introduce into our model the human behavior of the broker by using the following binary data:

- Dependent / Independent.
- Calm / Anxious.
- Intuitive / Analytic.
- Expert / Non-expert.

For this behavior, we calculate other variables such as risk factor and delay in the operations performed by a trader. This behavior will be the basis of their particular conduct when buying and selling shares, and it will determine the price they are willing to pay or receive for them. Added to these features of the trader, we have included in our system a factor of irrational luck, which is none other than the fact that each broker (particularly those with no real share price) has a hunch about the real price, i.e., an initial guess on its value, which, according to his or her personality, will guide him more or less strongly.
With all these ingredients, we have built a simulation model that complies with the bases of the commented paper and is capable of simulating the behavior of a human broker. This is done through a technical analysis that provides information on fluctuations in the market, and it is influenced by the psychological character of their profile. Fig. 4 shows the formalization of this decision in the framework of the model.
Since we need to compare the results with the results obtained in (Veiga & Vorsatz, 2008), it is necessary to define the same experimental framework or one that is as close as possible to (Veiga & Vorsatz, 2008)'s paper. That means:

1. We have a total of 24 traders, which we divide into two groups of 12.
2. Each group will be subjected to 10 rounds of 5 minutes each in a double auction market.
3. The first 2 rounds are for testing purposes. The study starts from the 3rd, and consequently, we discard the first two rounds.
4. At the beginning of round 3, 3 selected traders know the actual price of the share: 220 ECU or 100 ECU, also chosen randomly. At the end of the experiment, all traders have been informed a total of 2 times.
5. At the beginning of round 3, the robot[5] is added to the group (representing the manipulator agent), making the total of traders in the group 13.
6. Each trader has ECU 6000 and 2 shares at the beginning of each round.
7. The robot cannot be one of the reported traders.
8. The robot is involved in the second round from the 25th second until 50 seconds remain.
9. The robot's goal is to buy 10 shares and sell them later.
10. When the robot purchases a share, it increases the value of the higher takeover offer by a maximum of 10 ECU.
11. The robot gives a discount of up to 8 ECU below the value of the lowest offer.
12. The delay between the actions of the robot is 3.

The simplification assumptions for the experimental framework we are using are:

1. We have just 8 rounds, as in our case the human traders are simulated by computer and do not need any prior training. For that, we discard the two initial rounds since they are not considered in the original experiment (Veiga & Vorsatz, 2008) and because the intention is to familiarize the participants with the system. In our case, these two rounds are obviously not needed.
2. We have one group of 13 traders; the last of these, the robot manipulator, can be eliminated by reconfiguring the system variables so they are not allowed to participate in the simulation.
3. The selection of traders who know the value is random, but neither the actual price of the action nor the delay and offer of the robot is random. This is so because we are interested in studying further means for having control over system settings and thus simulating different environments. To do this for the first round, we select the first, second and third traders; for the second round the fourth, fifth and sixth, and so on. In the 8th round we select the tenth, eleventh and twelfth traders.

As we said previously, the trader agent can behave as a robot or a human, depending on the configuration.

[5] The trading robot (which represents the manipulator agent), is a computer program that attempts to distort the market and, thus, tries to profit from their operations. In our approach, this is a simple algorithm that buys 10 shares and sells them later.

Therefore, we are not establishing a separate agent to represent the robot. This simplifies the parameterizations of the several simulations that must be performed in order to represent the paper's scenario. All agents, including Company, consider the time to mark each round and the end of the simulation. In the case where a round ends, the values are reset while in the case where the simulation ends, we send each agent to a well state: *Terminate*. The different scenarios of our experiment are summarized in **Table 1**. Regarding the psychological characteristics of the traders a huge amount of combinations can be defined here. In our experiment we fix theses psychological traits as summarized in **Table 2**.

**Table 1**.The different scenarios of our experiment. MXXX represents those scenarios with a manipulative agent.

| Experiment | Manipulator agent | Price |
|---|---|---|
| B100 | No | 100 |
| M100 | Yes | 100 |
| B220 | No | 220 |
| M220 | Yes | 220 |

**Table 2**. Psychological characteristics of the traders of the experiment. The selection of these features for the brokers is based on the assumptions of our model and a combination of the features.

| | Trader | | | | | | | | | | | |
|---|---|---|---|---|---|---|---|---|---|---|---|---|
| | 1 | 2 | 3 | 4 | 5 | 6 | 7 | 8 | 9 | 10 | 11 | 12 |
| Experts | X | X | X | X | X | X | X | X | | | | |
| Analytic | X | X | X | X | | | | | X | X | | X |
| Calm | X | X | | | X | X | | | X | X | X | |
| Dependent | X | | X | | | X | | X | X | | X | X |

## 5. Formalizing and implementing the model

Following the approach to define a conceptual model that will be compressible by all the stakeholders, and to simplify the Validation and Verification processes we use Specification and Description Language (SDL). Specifically, in the world of simulation, different formal languages can be used, like SysML (OMG SysML, 2010), DEVS (Zeigler, Praehofer, & Kim, 2000), Petri Nets (Petri, 1962) (Silva Suárez, 1985) or Specification and Description Language (SDL) (ITU-T, 2012). We use in this work SDL, mainly because of its modular structure, graphical capabilities, lack of ambiguity and because it is a standard language, standardized under the Z.100 recommendation of the ITU-T (ITU-T, 2012) (Doldi L. , 2003) (Doldi L. , 2001). Also, it can be understood by several tools and can be used to implement the model, like (Fonseca, 2012), (PragmaDev SARL, 2012), (Sandrila Ltd, 2009), (IBM, 2009) among others.
The model is defined using Microsoft Visio diagrams with the SanDriLa plugin (Sandrila Ltd, 2009). Thanks to this graphical representation of the model, we can detail the behavior of the different agents that compose the model without the need to understand or explain anything related to its final implementation. The tool used to perform the execution form the model SDL representation is SDLPS, and the model can be seen online at https://sdlps.com/ or in the permanent repository REF. This was really useful for the specialist who helped us define the psychological behavior of the traders. In Fig.3, the system diagram for the model is shown (main page of SimTrader.vsd file).
+++

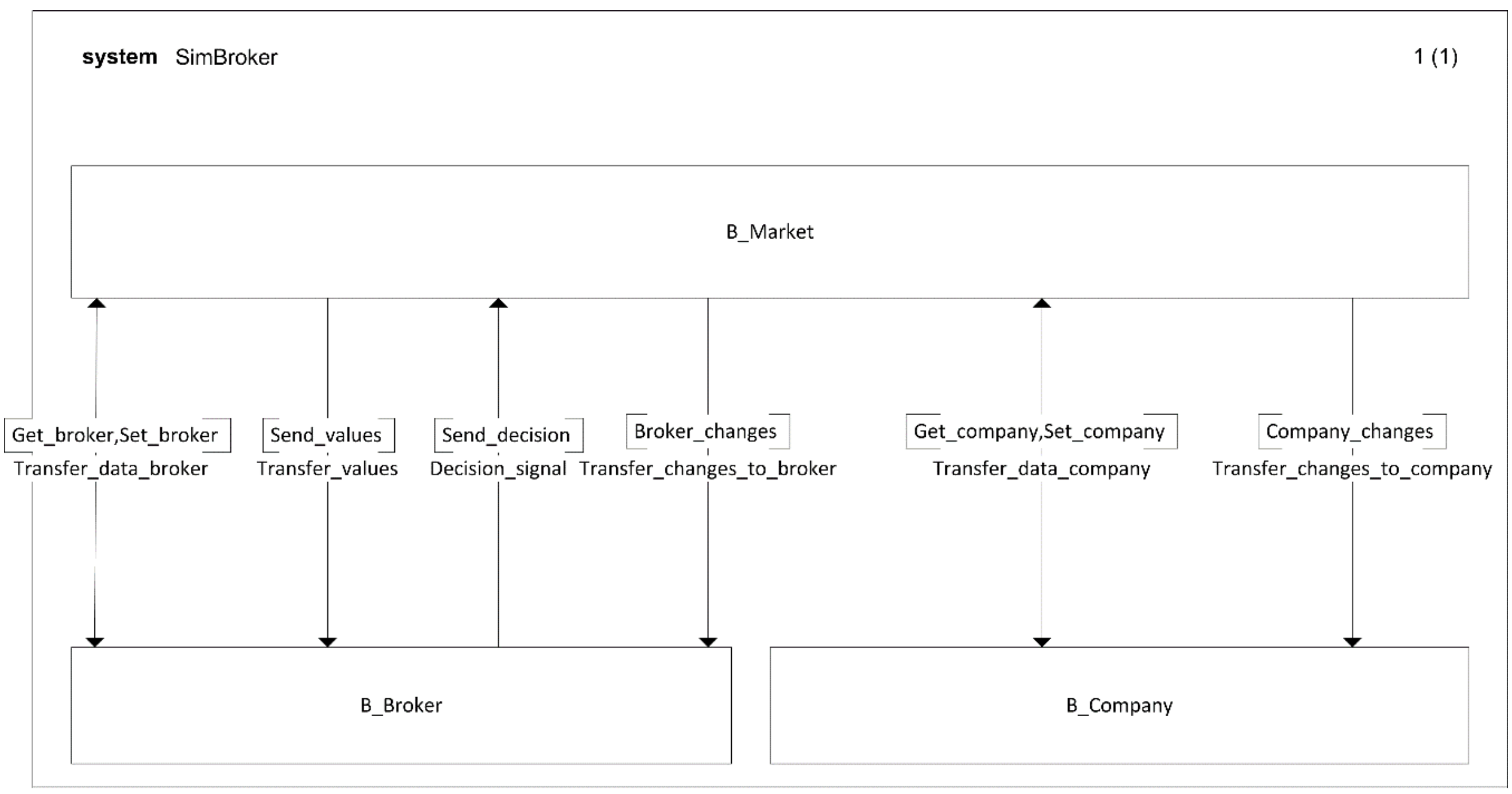

Fig.3. System diagram for the model. This first diagram shows: the market (B_Market) that defines the interchanges between the trades and the company; the brokers (B_Broker) that define the complete family of 12 brokers, plus the robot; and the company (B_Company) that defines the price of the asset. The entire model is represented using SDL diagrams. This helps in the validation of the model, since personnel without formation in computer science can ensure that the model behaves as they expect.

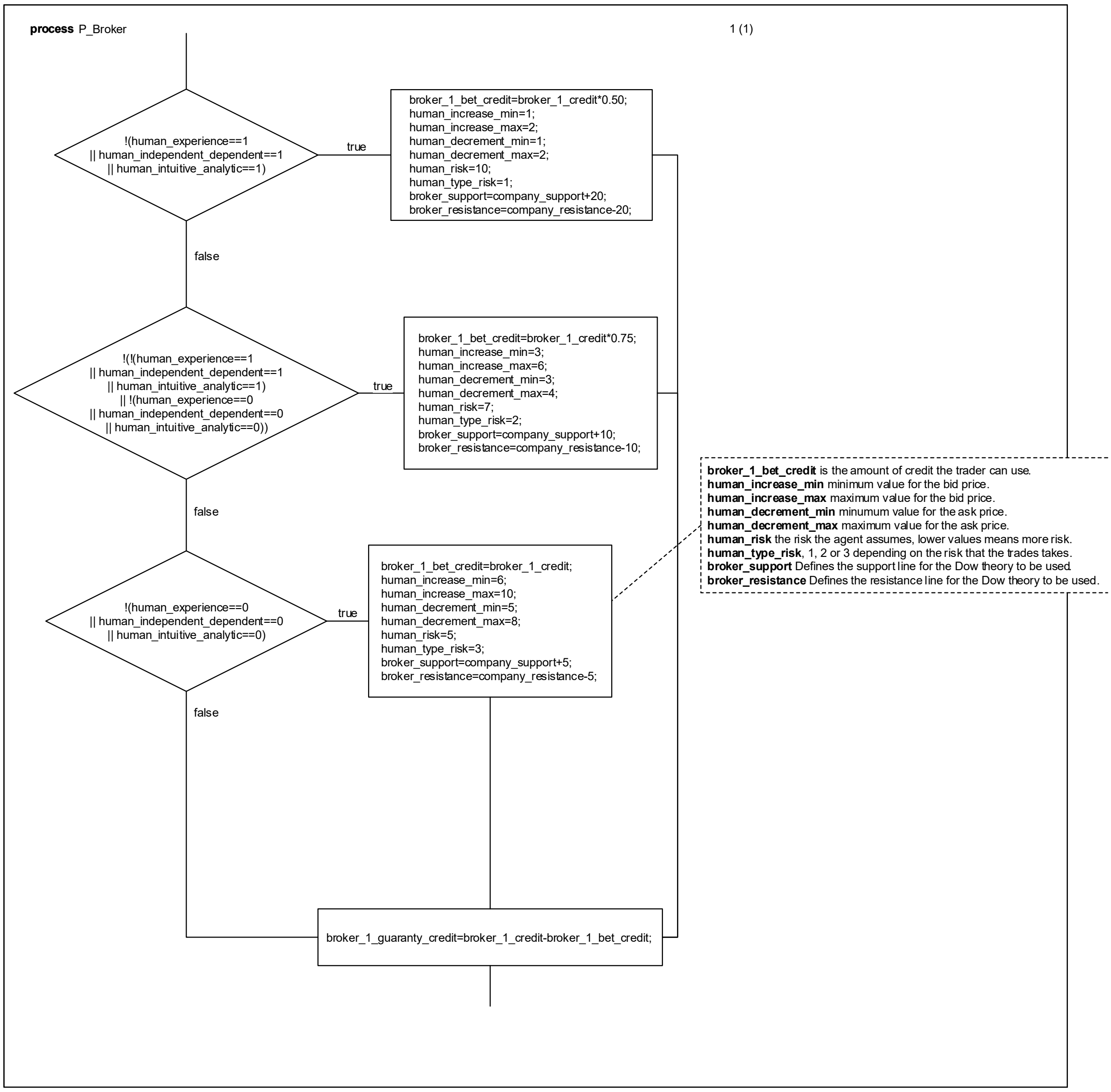

Fig. 4. SDL representation of the broker-agent risk definition. The three different profiles for risk are represented here. The main traits that categorize these profiles are experience, dependency and being analytic; depending on this we obtain a low risk profile where, as an example, the amount of credit to be used will not be greater than 50% of the existing credit. The complete formalization can be reviewed on the annexes 13.1.

## 5.1. Implementation of the model

As we mention, once we have the formalization of the model, it is not necessary to implement the simulator, since in our case we are using SDLPS (Fonseca i Casas & Casanovas, Towards a SDL-DEVS Simulator, 2011) (Fonseca i Casas, SDL distributed simulator, 2008) (Fonseca, 2012), a distributed simulator that understands Specification and Description language. Fig.5 shows the SDLPS local environment with the model loaded.

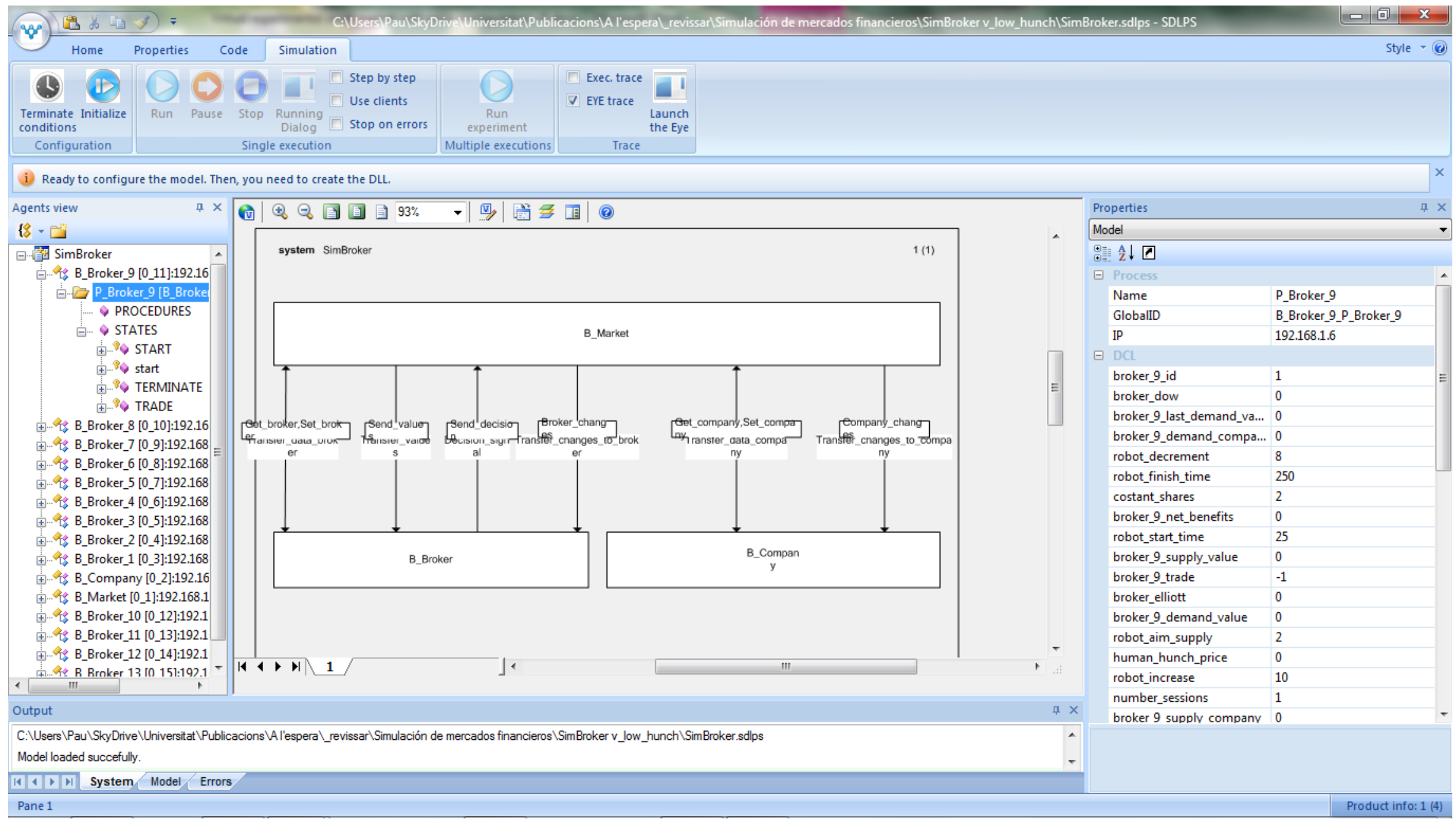

Fig.5. SDLS simulation environment. In the left window we have a detailed description of all the agents that compose the model. In our case, there are 13 brokers (traders). In the main window there is a representation of the SDL diagram that details the behavior of the selected agent. In the right window there is a detailed view of the declarations (DCL) that define the diagram variables. Validation of the modeling assumptions

### 5.2. Validation and verification of the model

We are not going to discuss the different aspects related to the verification of the model (the correctness of the implementation of the model). Instead, we will merely explain the formal representation of the model. Since the implementation is based on the representation of this model in SDL (thanks to the use of SDLPS, it is not necessary to perform an implementation once the model is specified in SDL), we can detect some implementation errors easily. We perform an Operational Validation, comparing the results obtained from the simulation model with those obtained in the experiment conducted (Veiga & Vorsatz, 2008). Also, we use the model diagrams to perform a Conceptual Validation of the model in collaboration with the experts who help us in defining trader behavior. One of the main concerns of this model is that it is very sensitive to modifications on the traits of the agents. This implies that slight modifications lead to completely new solutions. Here is where the formal representation of the model helps. With it, we can take the data obtained in the Veiga and Vorsatz experiment and compare it with the diagrams that lead to this model-specific behavior.

The process we follow to validate the model is represented in the next picture, following the approach presented in (Sargent, 2007). Red represents the processes that are simplified thanks to the methodology we follow. The formal representation of the model allows us to perform a Conceptual Model Validation, while the automatic implementation of the formal model through SDLPS simplifies the Verification process.

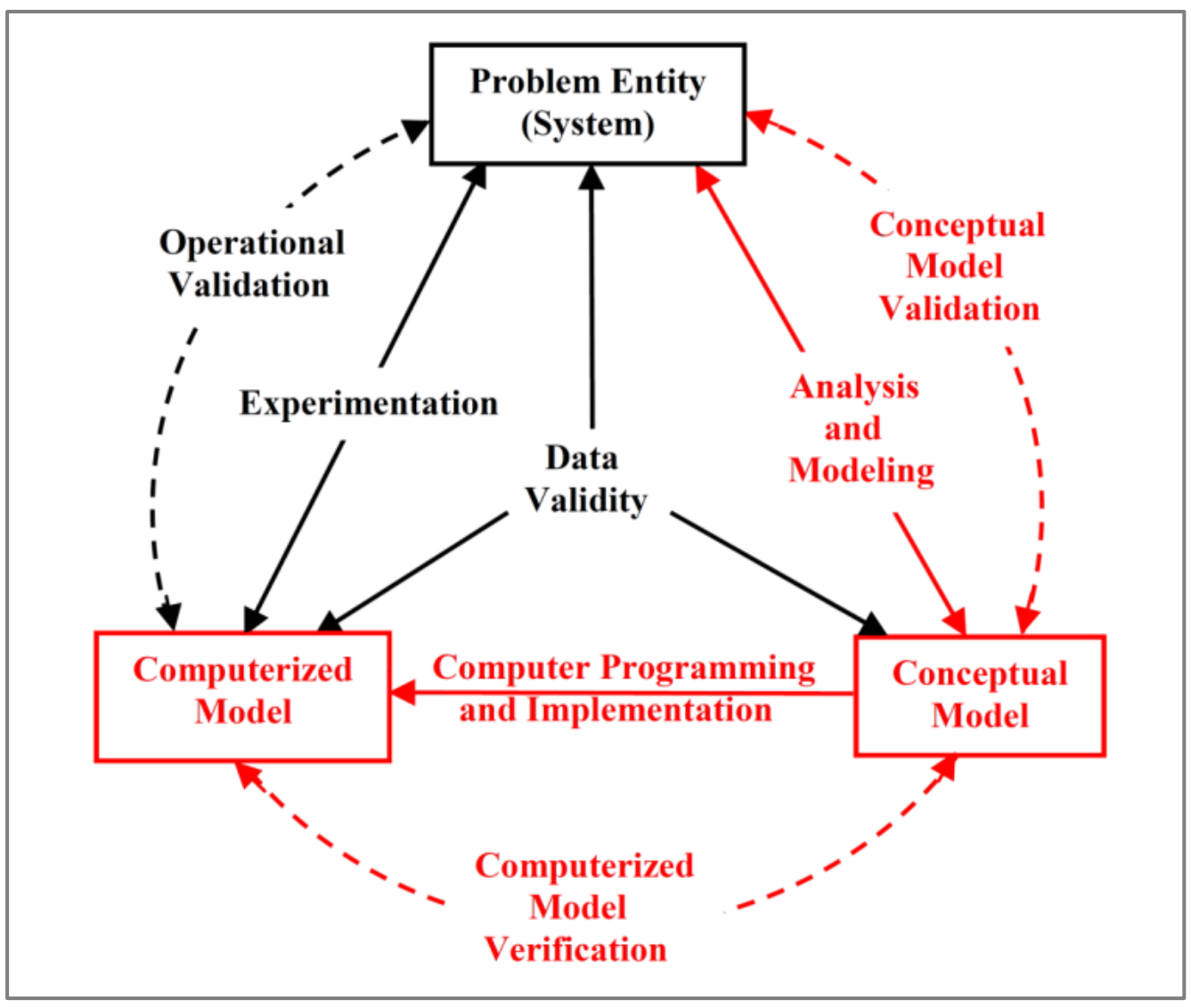

Fig.6. In red we can see the areas of the VV process that are simplified thanks to this approach.

One interesting feature analyzed through the course of the Conceptual Model Validation, which determines the correct behavior of the agents, was the management of the hunches. The first models consider that the agents believe strongly in their hunches. When we perform the Operational Validation, these preliminary models give results that are not correct when compared with the results obtained by (Veiga & Vorsatz, 2008). Hence, it is necessary to reconsider the effects of the hunches on the model. This led us to rewrite some assumptions on the SDL diagrams, obtaining a more accurate model that now achieves Operational Validation. We also review the definition of the model based on SDL diagrams and perform a new Conceptual Model Validation. This allows us to conclude that the assumptions are coherent with the theoretical background we use to define the physiological traits of our agents. To review the data of the preliminary analysis annex 13.2 can be consulted.

### 5.3. Statistical analysis of the traits

Table 3. Effect of the “expert” psychological feature for no informed traders. The two distributions have not so different medians (true location shift is not equal to 0).

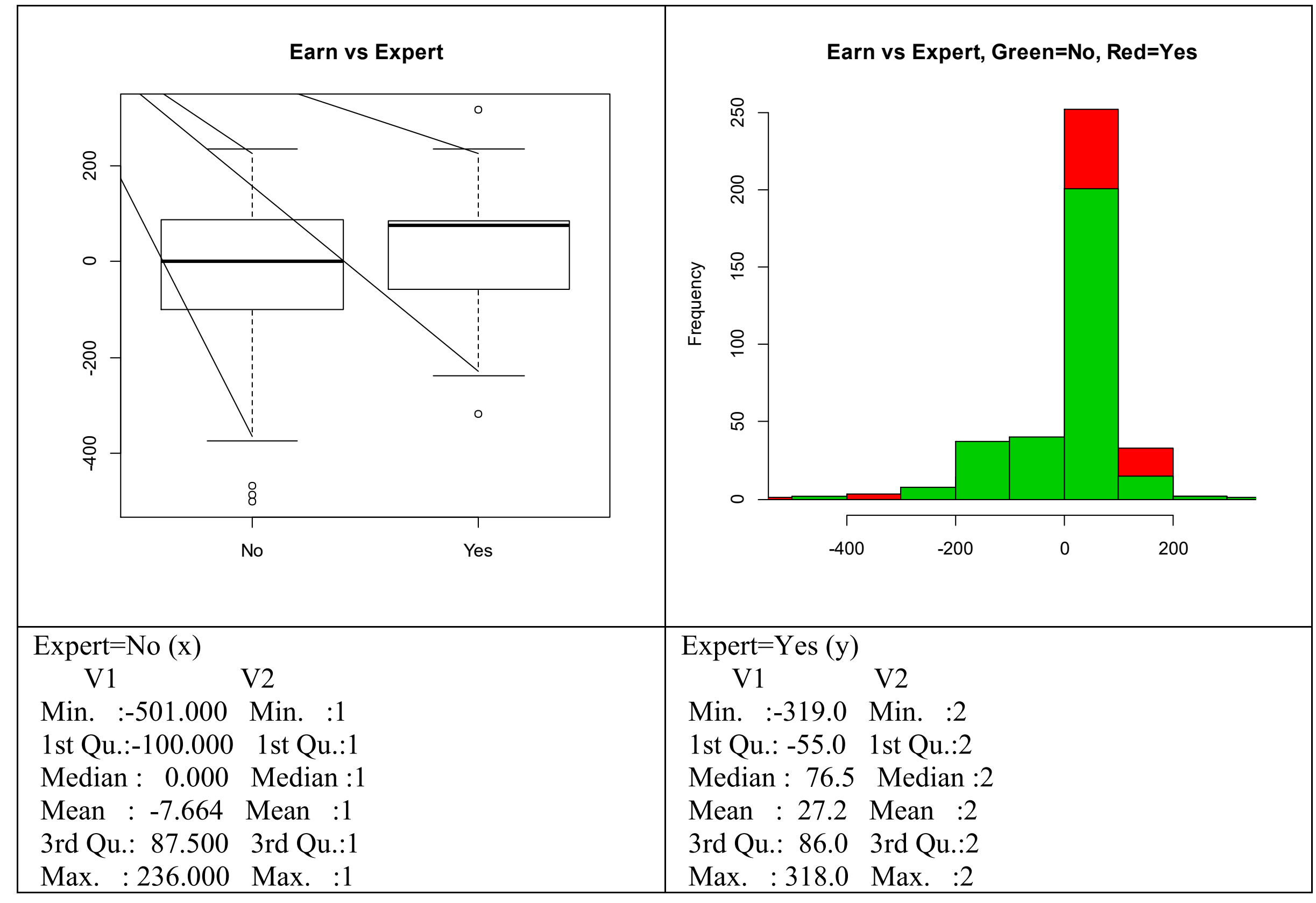

| Expert=No (x) | | Expert=Yes (y) | |
|---|---|---|---|
| V1 | V2 | V1 | V2 |
| Min. :-501.000 | Min. :1 | Min. :-319.0 | Min. :2 |
| 1st Qu.:-100.000 | 1st Qu.:1 | 1st Qu.: -55.0 | 1st Qu.:2 |
| Median : 0.000 | Median :1 | Median : 76.5 | Median :2 |
| Mean : -7.664 | Mean :1 | Mean : 27.2 | Mean :2 |
| 3rd Qu.: 87.500 | 3rd Qu.:1 | 3rd Qu.: 86.0 | 3rd Qu.:2 |
| Max. : 236.000 | Max. :1 | Max. : 318.0 | Max. :2 |

```
> wilcox.test(Earn ~ Expert, alternative="two.sided", data=Traders_no_Informed)

Wilcoxon rank sum test with continuity correction

data:  Earn by Expert
W = 11016, p-value = 0.05808
alternative hypothesis: true location shift is not equal to 0
```

Table 4. Effect of the "analytic" psychological feature for no informed traders, both distributions have different medians (true location shift is not equal to 0).

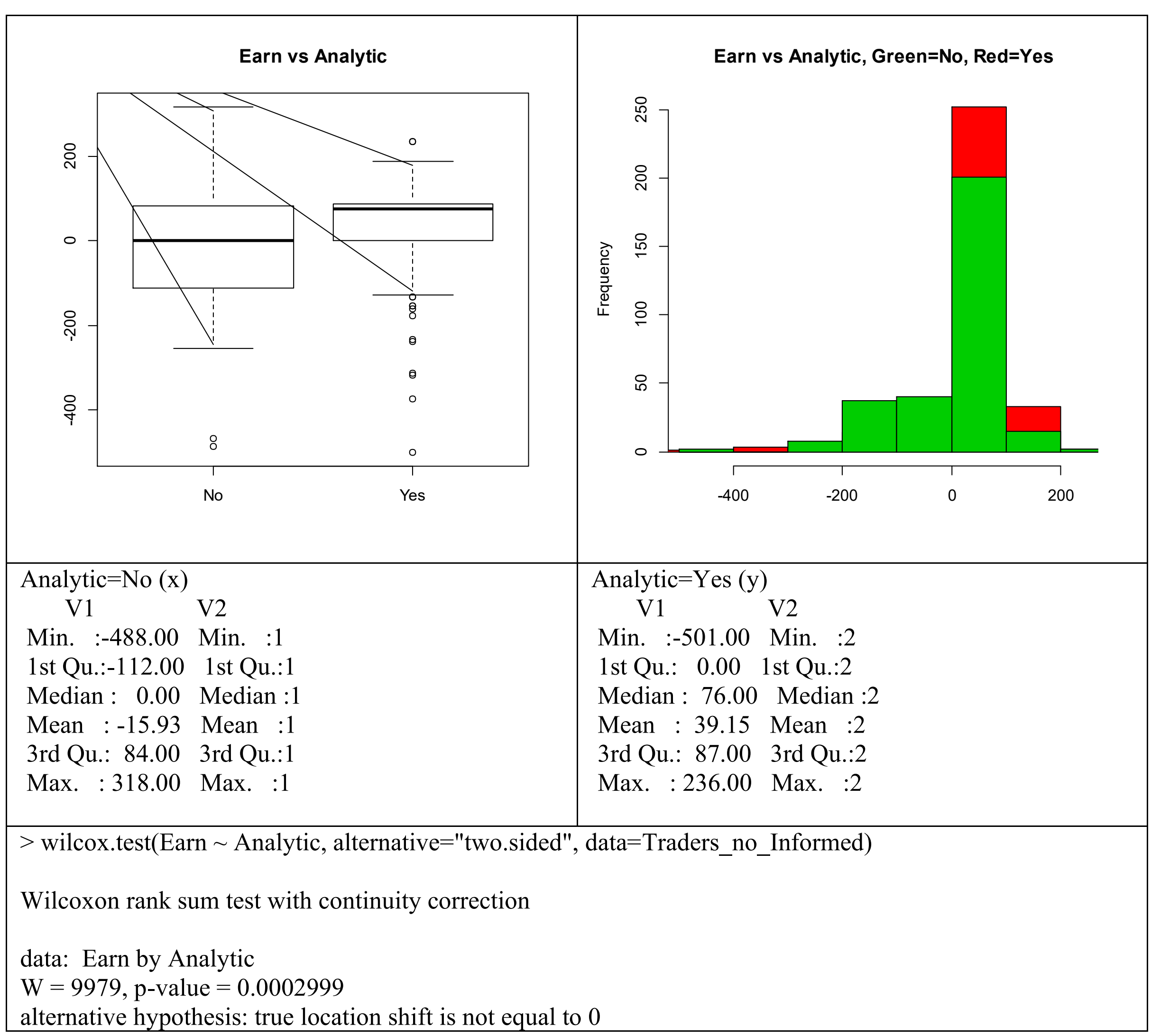

| Analytic=No (x) | Analytic=Yes (y) |
|---|---|
| V1 V2 | V1 V2 |
| Min. :-488.00 Min. :1 | Min. :-501.00 Min. :2 |
| 1st Qu.:-112.00 1st Qu.:1 | 1st Qu.: 0.00 1st Qu.:2 |
| Median : 0.00 Median :1 | Median : 76.00 Median :2 |
| Mean : -15.93 Mean :1 | Mean : 39.15 Mean :2 |
| 3rd Qu.: 84.00 3rd Qu.:1 | 3rd Qu.: 87.00 3rd Qu.:2 |
| Max. : 318.00 Max. :1 | Max. : 236.00 Max. :2 |

```
> wilcox.test(Earn ~ Analytic, alternative="two.sided", data=Traders_no_Informed)

Wilcoxon rank sum test with continuity correction

data:  Earn by Analytic
W = 9979, p-value = 0.0002999
alternative hypothesis: true location shift is not equal to 0
```

Table 5. Effect of the "calm" psychological feature for no informed traders. The two distributions have not so different medians (true location shift is not equal to 0).

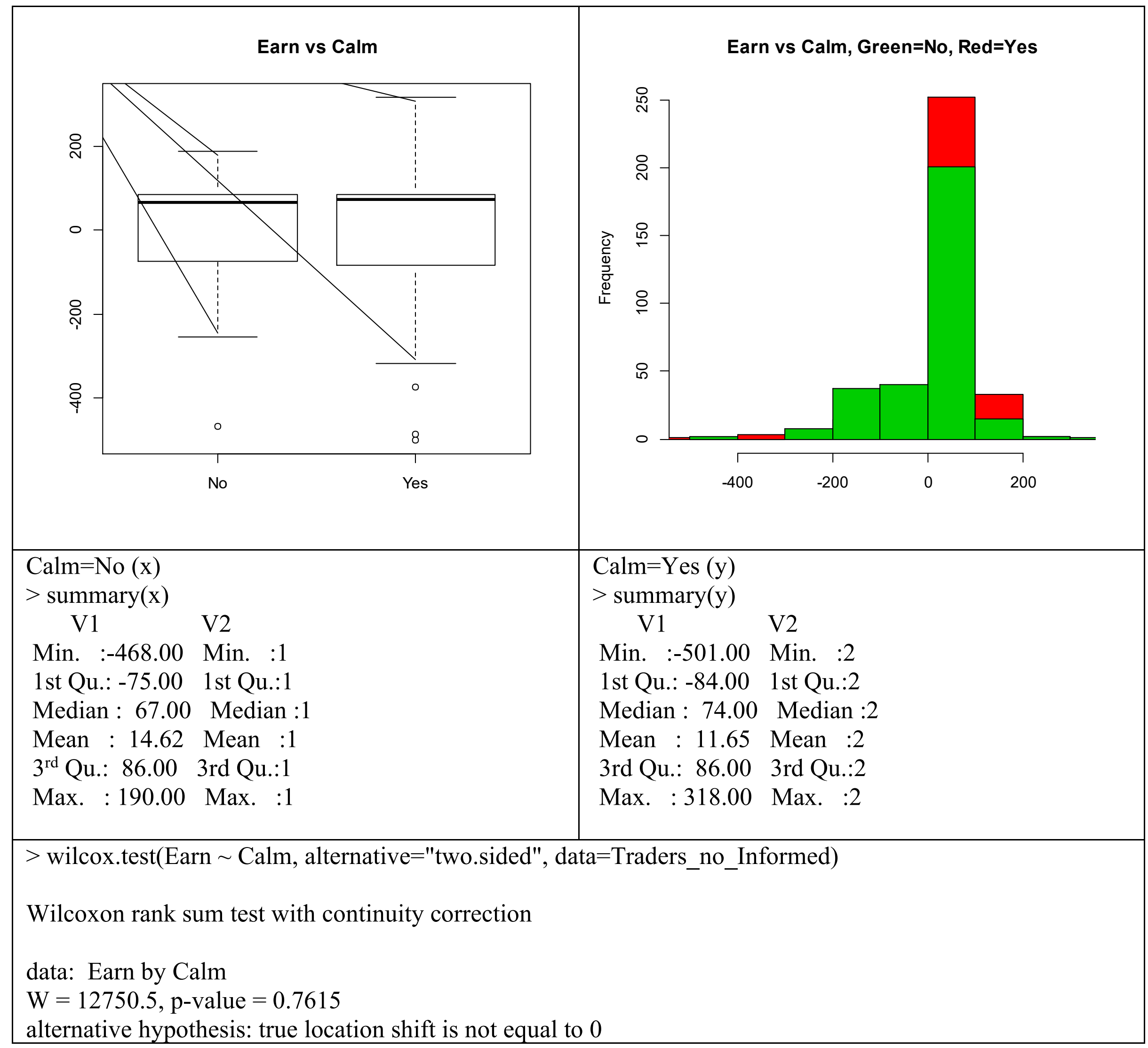

```
Calm=No (x)
> summary(x)
       V1               V2
 Min.   :-468.00   Min.   :1
 1st Qu.: -75.00   1st Qu.:1
 Median :  67.00   Median :1
 Mean   :  14.62   Mean   :1
 3rd Qu.:  86.00   3rd Qu.:1
 Max.   : 190.00   Max.   :1
```

```
Calm=Yes (y)
> summary(y)
       V1               V2
 Min.   :-501.00   Min.   :2
 1st Qu.: -84.00   1st Qu.:2
 Median :  74.00   Median :2
 Mean   :  11.65   Mean   :2
 3rd Qu.:  86.00   3rd Qu.:2
 Max.   : 318.00   Max.   :2
```

```
> wilcox.test(Earn ~ Calm, alternative="two.sided", data=Traders_no_Informed)

Wilcoxon rank sum test with continuity correction

data:  Earn by Calm
W = 12750.5, p-value = 0.7615
alternative hypothesis: true location shift is not equal to 0
```

## 6. Results of the model

We show the evolution of the mean asset values for all the rounds in each scenario. In this case, the presence of the manipulative agent causes the values to grow. However, the models with a manipulative agent always obtain larger values than in the other scenarios. Fig.7 shows the average calculated value of the company of all simulated rounds for the different cases: the real value contribution of ECU 100, without the presence of the manipulative agent (B100) and with the presence of the manipulative agent (M100); and the real value contribution of 220 ECU, without the presence of the agent handler (B220) and with the presence of the manipulative agent (M220). We can see that there is greater variation in share price if the manipulative agent is present in the simulation. In these cases, stocks grow faster when the manipulative agent starts its actions within 25 units of time, as seen in the experiment.

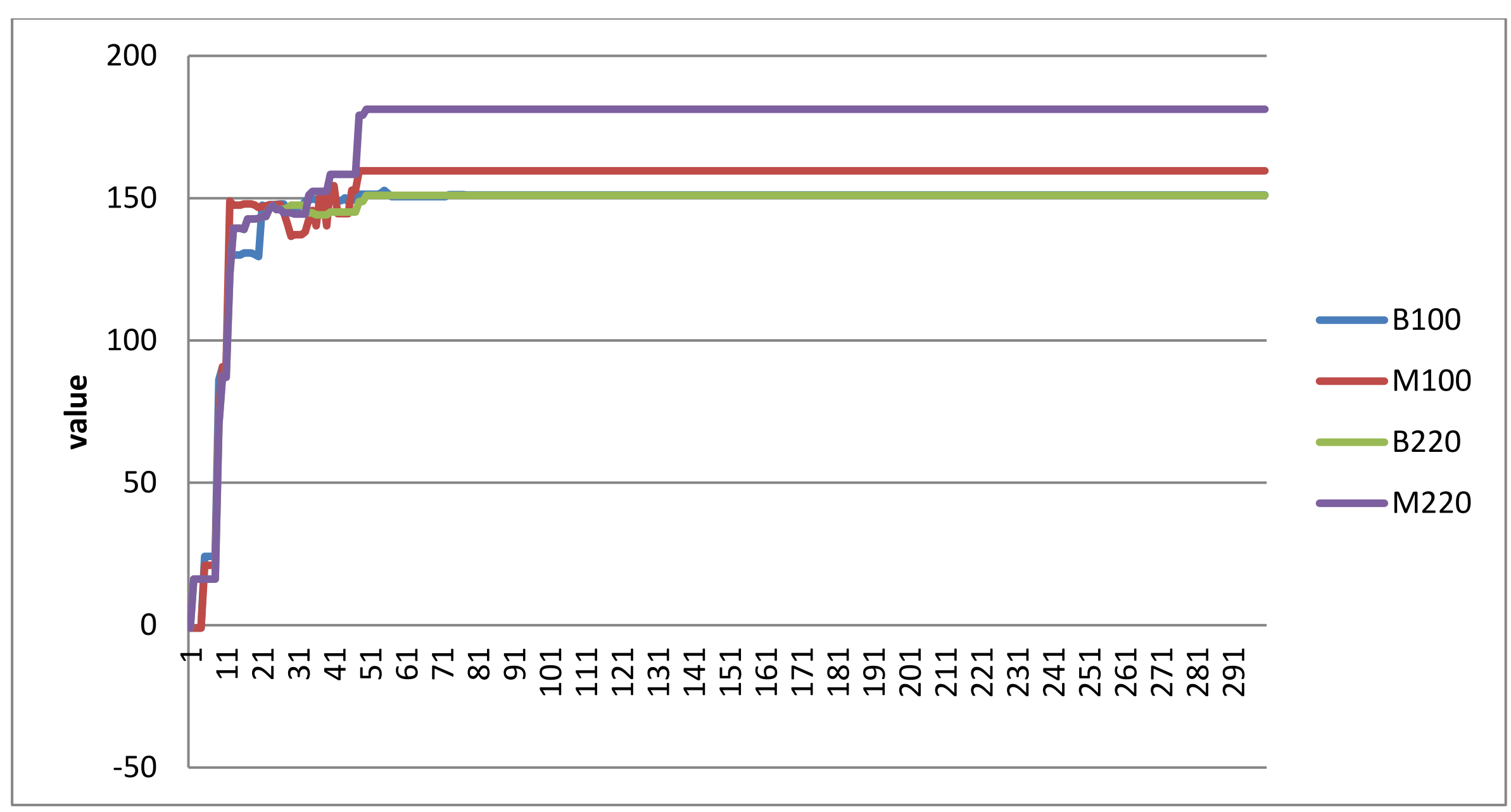

Fig.7. Means of the values obtained in each one of the different experiments during the simulation. On the abscise axis we represent time, while on the ordinate axis the value of the asset.

In our model the agents do not believe so much in their hunches and therefore observe the market before executing a decision, being more influenced by manipulative agents. This implies that, in the scenarios where a manipulative agent is present, the final value is higher than in the other scenarios (M100>B100 and M220>B220). Again, once the evolution of the market value action is seen and understood, we find the average of the profits for each trader in each of the different scenarios. **Table 6**, **Table 7**, **Table 8** and **Table 9** show the results of the experiments, depending on the parameterization used.

**Table 6**. No manipulation agent and value of 100 (experiment B100). This table shows the earnings (or losses) for each trader in the 8 rounds of the B100 experiment.

| | Rounds | | | | | | | | |
|---|---|---|---|---|---|---|---|---|---|
| Trader | 1 | 2 | 3 | 4 | 5 | 6 | 7 | 8 | Mean |
| 1 | 72 | -319 | -161 | -84 | -92 | -75 | -92 | 135 | -77 |
| 2 | 86 | 150 | 64 | 76 | 76 | 154 | 86 | 135 | 103,375 |
| 3 | 79 | -239 | -128 | -52 | 77 | -75 | -132 | 135 | -41,875 |
| 4 | 78 | 158 | 64 | 76 | 76 | 133 | 78 | 135 | 99,75 |
| 5 | -232 | 150 | -193 | -231 | -196 | 150 | -238 | 135 | -81,875 |
| 6 | 82 | 160 | 79 | 80 | 80 | 149 | 86 | 70 | 98,25 |
| 7 | -82 | 0 | 64 | -40 | -76 | 143 | -40 | 72 | 5,125 |
| 8 | 94 | 154 | 65 | 84 | 84 | 143 | 94 | 88 | 100,75 |
| 9 | -178 | -374 | 61 | -116 | -116 | -501 | 82 | 71 | -133,875 |
| 10 | 78 | 149 | 71 | 76 | 76 | 128 | 78 | 135 | 98,875 |
| 11 | -125 | -153 | -64 | 87 | -77 | -488 | -50 | 135 | -91,875 |
| 12 | 92 | 164 | 78 | 88 | 88 | 139 | 92 | 135 | 109,5 |

**Table 7**. Manipulative agent and value of 100 (experiment M100). This table shows earnings (or losses) for each trader in the 8 rounds of the M100 experiment.

| | Rounds | | | | | | | | |
|---|---|---|---|---|---|---|---|---|---|
| Trader | 1 | 2 | 3 | 4 | 5 | 6 | 7 | 8 | Mean |
| 1 | 72 | -235 | -118 | -84 | -40 | -75 | -46 | 57 | -58,625 |
| 2 | 86 | 150 | 64 | 76 | 76 | 154 | 86 | 76 | 96 |
| 3 | 79 | -155 | -64 | 0 | 77 | -75 | -86 | -128 | -44 |
| 4 | 78 | 144 | 64 | 76 | 76 | 0 | 78 | 76 | 74 |
| 5 | -194 | 150 | -164 | -152 | -196 | 150 | -200 | -210 | -102 |
| 6 | 82 | 160 | 79 | 80 | 80 | 80 | 86 | 80 | 90,875 |
| 7 | -82 | 0 | 64 | -40 | -76 | 74 | -40 | -76 | -22 |
| 8 | 94 | 154 | 36 | 84 | 84 | 79 | 94 | 84 | 88,625 |
| 9 | -132 | -155 | 32 | -116 | -116 | -314 | 82 | -116 | -104,375 |
| 10 | 78 | 144 | 71 | 76 | 76 | 0 | 78 | 53 | 72 |
| 11 | -125 | -79 | -64 | 97 | -77 | -232 | -50 | 128 | -50,25 |
| 12 | 98 | 190 | 104 | 100 | 130 | 159 | 101 | 80 | 120,25 |
| 13 | -90 | -468 | -104 | -153 | -94 | 0 | -139 | -104 | -144 |

**Table 8**. No manipulation agent and value of 220 (experiment B220). This table shows the earnings (or losses) for each trader in the 8 rounds of the B220 experiment.

| | Rounds | | | | | | | | |
|---|---|---|---|---|---|---|---|---|---|
| Trader | 1 | 2 | 3 | 4 | 5 | 6 | 7 | 8 | Mean |
| 1 | 74 | 74 | 236 | 0 | 74 | 74 | 74 | -83 | 65,375 |
| 2 | 0 | 86 | 69 | 86 | 0 | 86 | 86 | 86 | 62,375 |
| 3 | 57 | 66 | 67 | 0 | 66 | 0 | 66 | 74 | 49,5 |
| 4 | 86 | 0 | 77 | 86 | 0 | 86 | 86 | 78 | 62,375 |
| 5 | 160 | 156 | 176 | 160 | 156 | 160 | 236 | 318 | 190,25 |
| 6 | 90 | -46 | 81 | 90 | 86 | -46 | 94 | 86 | 54,375 |
| 7 | 80 | 80 | -80 | 80 | 80 | 80 | 0 | 80 | 50 |
| 8 | -146 | -146 | 251 | -146 | -146 | -146 | 74 | -146 | -68,875 |
| 9 | 74 | 74 | 0 | 74 | 74 | 148 | 0 | 236 | 85 |
| 10 | 86 | 132 | 65 | 0 | 86 | 87 | 0 | 78 | 66,75 |
| 11 | -146 | -100 | -74 | -146 | -100 | -97 | -101 | -87 | -106,375 |
| 12 | 109 | 104 | 92 | 0 | 104 | 92 | 105 | 0 | 75,75 |

**Table 9**. Manipulation agent and value of 220 (experiment M220). This table shows the earnings (or losses) for each trader in the 8 rounds of the M220 experiment.

| | Rounds | | | | | | | | |
|---|---|---|---|---|---|---|---|---|---|
| Trader | 1 | 2 | 3 | 4 | 5 | 6 | 7 | 8 | Mean |
| 1 | 0 | 0 | 161 | 0 | 0 | 0 | 0 | -83 | 9,75 |
| 2 | 0 | 86 | 69 | 86 | 0 | 86 | 102 | 86 | 64,375 |
| 3 | 0 | 0 | 0 | 0 | 0 | 0 | 0 | 74 | 9,25 |
| 4 | 86 | 0 | 69 | 86 | 0 | 86 | 86 | 86 | 62,375 |
| 5 | 160 | 156 | 176 | 160 | 156 | 160 | 156 | 236 | 170 |
| 6 | 90 | 0 | 81 | 90 | 90 | -46 | 94 | 86 | 60,625 |
| 7 | 80 | 80 | -80 | 80 | 80 | 80 | 0 | 80 | 50 |
| 8 | -146 | -146 | 251 | -146 | -146 | -146 | 0 | -146 | -78,125 |

| 9 | 74 | 0 | 0 | 74 | 0 | 74 | 0 | 154 | 47 |
|---|---|---|---|---|---|---|---|---|---|
| 10 | 86 | 86 | 65 | 0 | 86 | 87 | 0 | 78 | 61 |
| 11 | -146 | -100 | -74 | -146 | -100 | -97 | -101 | -87 | -106,375 |
| 12 | 177 | 156 | 124 | 0 | 156 | 112 | 157 | 0 | 110,25 |
| 13 | -177 | -198 | -122 | 0 | -202 | -112 | -254 | -84 | -143,625 |

Next we are going to perform an individual analysis for each one of the traders. B100, B220, M100 and B220 in **Table 10** show the average earnings of the trader in each experiment for the 8 rounds. Each one shows the mean of trader profits for all the experiments and rounds. The columns related to the "manipulative agent" represent the mean of the values for the experiments, with the presence or absence of the manipulative agent (M100 and M220 are represented in the "Yes" column). The "Influence?" column shows if the manipulative agent has any effect on trader behavior. The "Hunch" column shows if the trader follows a hunch (1 indicates he or she believe the value is about 100, and 0 means a value of 220). Finally, the "Earnings?" column shows if the agent has profits at the end of the whole simulation.

**Table 10**.This table details earnings for each of the traders. _The "Influent?" column is true if the earnings with the manipulative agent are greater than the earning without the manipulative agent.

| | | | | | Average earnings | | | | | |
|---|---|---|---|---|---|---|---|---|---|---|
| | | | | | | Manipulative agent | | | | |
| Trader | B100 | B220 | M100 | M220 | All | No | Yes | Influent? | Hunch | Earnings? |
| 1 | -77 | 65,37 | -58,625 | 9,75 | -15,125 | -5,8125 | -24,4375 | 0 | 0 | 0 |
| 2 | 103,37 | 62,37 | 96 | 64,375 | 81,53125 | 82,875 | 80,1875 | 0 | 1 | 1 |
| 3 | -41,87 | 49,5 | -44 | 9,25 | -6,78125 | 3,8125 | -17,375 | 0 | 0 | 0 |
| 4 | 99,75 | 62,375 | 74 | 62,375 | 74,625 | 81,0625 | 68,1875 | 0 | 1 | 1 |
| 5 | -81,875 | 190,25 | -102 | 170 | 44,09375 | 54,1875 | 34 | 0 | 0 | 1 |
| 6 | 98,25 | 54,37 | 90,87 | 60,625 | 76,03125 | 76,3125 | 75,75 | 0 | 1 | 1 |
| 7 | 5,125 | 50 | -22 | 50 | 20,78125 | 27,5625 | 14 | 0 | 0 | 1 |
| 8 | 100,75 | -68,875 | 88,625 | -78,125 | 10,59375 | 15,9375 | 5,25 | 0 | 1 | 1 |
| 9 | -133,87 | 85 | -104,37 | 47 | -26,5625 | -24,4375 | -28,6875 | 0 | 0 | 0 |
| 10 | 98,87 | 66,75 | 72 | 61 | 74,65625 | 82,8125 | 66,5 | 0 | 1 | 1 |
| 11 | -91,87 | -106,37 | -50,25 | -106,375 | -88,71875 | -99,125 | -78,3125 | 1 | 0 | 0 |
| 12 | 109,5 | 75,75 | 120,25 | 75,75 | 103,9375 | 92,625 | 115,25 | 1 | 1 | 1 |

Finally, we obtain the earnings for the traders, depending on if they have been informed or not (see **Table 11**, **Table 12**, **Table 13** and **Table 14**).

**Table 11**. No manipulation agent and trader's profits with a value of 100 (experiment B100).

| B100 | Rounds | | | | | | | | |
|---|---|---|---|---|---|---|---|---|---|
| | 1 | 2 | 3 | 4 | 5 | 6 | 7 | 8 | Mean |
| Informed | 79 | 156 | 63,3 | 83,6 | 76,3 | 147,3 | 84,6 | 135 | 103,16 |
| Not informed | -19,3 | -46,8 | -19 | -20,7 | -22,9 | -44,2 | -21 | 111,1 | -10,35 |
| Informed earnings? | 1 | 1 | 1 | 1 | 1 | 1 | 1 | 1 | 1 |

**Table 12**. Manipulation agent and trader's profits with a value of 100 (experiment M100)

| M100 | Rounds | | | | | | | | |
|---|---|---|---|---|---|---|---|---|---|

| | 1 | 2 | 3 | 4 | 5 | 6 | 7 | 8 | Mean |
|---|---|---|---|---|---|---|---|---|---|
| Informed | 79 | 151,3 | 44 | 91 | 76,3 | 101,3 | 84,6 | 88,3 | 89,5 |
| Not informed | -19,3 | -45,4 | -13,2 | -22,9 | -22,9 | -30,4 | -21 | -26,5 | -25,2 |
| Informed earnings? | 1 | 1 | 1 | 1 | 1 | 1 | 1 | 1 | 1 |

**Table 13**. No manipulation agent and trader's profits with a value of 220 (experiment B220)

| B220 | Rounds | | | | | | | | |
|---|---|---|---|---|---|---|---|---|---|
| | 1 | 2 | 3 | 4 | 5 | 6 | 7 | 8 | Mean |
| Informed | 43,6 | 36,6 | 57 | -48,6 | 22 | 64,6 | 24,6 | -56,6 | 17,91 |
| Not informed | 39,3 | 37 | 78,9 | 43 | 41,4 | 33 | 64,6 | 89 | 53,27 |
| Informed earnings? | 1 | 0 | 0 | 0 | 0 | 1 | 0 | 0 | 0 |

**Table 14**. Manipulation agent and trader's profits with a value of 220 (experiment M220).

| M220 | Rounds | | | | | | | | |
|---|---|---|---|---|---|---|---|---|---|
| | 1 | 2 | 3 | 4 | 5 | 6 | 7 | 8 | Mean |
| Informed | 0 | 52 | 57 | -48,6 | 0 | 64,6 | 0 | -56,6 | 8,541 |
| Not informed | 28,4 | -3,6 | 54,9 | 43 | 12 | 9 | 24 | 65 | 29,08 |
| Informed earnings? | 0 | 1 | 1 | 0 | 0 | 1 | 0 | 0 | 0 |

## 7. Results analysis: comparing the results with the original experimental results

The primary objective of this paper is to reproduce the behavior of a human trader, detailing his or her psychological processes in order to understand the effects of his or her decisions on the final value of an asset. To analyze if the implemented behavior is accurate, we compare our results with the results obtained in (Veiga & Vorsatz, 2008). A good indicator may be the means of market value obtained in both experiments.

**Table 15**. The results obtained from our experiment compared to the experiment of (Veiga & Vorsatz, 2008).

| Experiment | Manipulator agent | Price | Mean Value | Veiga and Vorsatz Mean Value |
|---|---|---|---|---|
| B100 | No | 100 | 151,12 | 132 |
| M100 | Yes | 100 | 159,62 | 183 |
| B220 | No | 220 | 151 | 218 |
| M220 | Yes | 220 | 181,25 | 217 |

With this new table, the next step is to verify again if the assumptions of (Veiga & Vorsatz, 2008) are in accordance with our results. Note that it is quite difficult in this kind of experiment to obtain a value that is close to the experimental value obtained in (Veiga & Vorsatz, 2008). We want the results obtained in the simulation model to approximate those obtained in the Veiga and Vorsatz experiment. Then, in order to try to understand the trader's behavior, we analyze if the better values in the virtual experiments are coherent with the better values in the real experiment.

### 7.1. Results of assumptions 1: price of the asset.

According to (Veiga & Vorsatz, 2008), this assumption is: "If the fundamental value of the asset is 100, the last contract price in the Manipulation Treatment is higher than the one in the Benchmark Treatment. If the fundamental value of the asset is 220, the contract price converges in both treatments to the fundamental value." In short, the assumption will be true if the manipulative agent can influence the price; the result is that this assumption is true if the estimated price for a given round is significantly higher when the manipulative agent is present. In (Veiga & Vorsatz, 2008), this assumption is **true**.
In our case this assumption is also **true**; the manipulative agent has an impact in both models, but a great impact when the value is 220.

Table 16. Final average value.

| True value: 100 ECU | True value: 220 ECU |
|---|---|

| Non-manipulative agent | Manipulative agent | Non-manipulative agent | Manipulative agent |
|---|---|---|---|
| 151,125 | 159,625 | 151 | 181,25 |

**7.2. Results of assumption 2: earnings from traders.**

The assumption of (Veiga & Vorsatz, 2008) is: "If the fundamental value of the asset is 100, the average payoff difference between informed and uninformed traders in Manipulation Treatment is greater than the one in the Benchmark Treatment. If the fundamental value of the asset is 220, the average payoff difference between informed and uninformed traders is the same for both treatments." In short, this assumption is true if the difference between the earnings of informed and uninformed agents is greater in M100 than in B100 and this difference is the same for M220 and B220. In (Veiga & Vorsatz, 2008) this assumption is **true**.
To analyze this in our models, we use Table 11, Table 12, Table 13 and Table 14. In our case, the difference between informed and uninformed agents in M100 is larger than in B100, but the difference is very small (the difference in M100 is 114.7 and in B100 is 113.5). Also, through analyzing the difference between informed and uninformed agents for M220 and B220, we see that this difference is negative (-35.36 for B220 and -20.54 for M220). This is because --in this case-- the informed agents, although they obtain profits, do not take advantage of their knowledge. Since the differences are small, and since we are presenting a deterministic model, we cannot conclude that the experiment confirms this assumption.

**7.3. Results of assumption 3: return of the manipulator**

Following (Veiga & Vorsatz, 2008)'s assumption: "*The manipulation program is not profitable*." This is **true** in (Veiga & Vorsatz, 2008).
This assumption will be true if the strategy undertaken by the manipulative agent is profitable, or in other words, obtains profits. Just look, in this case, at the table of profits for the experiments with the manipulative agent (**Table 7** and **Table 9**). As we can see in our model, this assumption is **true**.

## 8. Concluding remarks

This paper shows a model that represents the behavior of a trader from the point of view of his psychological beliefs. To do this, we present first a formalization of this behavior using Specifications and Description Language (SDL). SDL helps us in the representation of the structure and the detailed behavior of the agents also helps in the implementation of the model and in its Conceptual Validation.
Regarding the results of the model, we would like to remark that the main objective is to develop a model that reproduces the experiment presented by Veiga and Vorstaz, focusing on the psychological representation of the traders' behavior using a detailed psychological representation.
In our model we see that our conclusions are coherent with the results obtained from the Veiga and Vorstaz paper. Thus, we can suppose that the behavior of the trader can be represented by the psychological profiles we define. Based on the psychological assumptions of the model, we can extract some interesting conclusions that have some similarities with common sense. The main psychological variables we are studying are, for each trader:

1. Emotional adjustment - Resistance to stress (calm - anxious)
2. Social dependency (independent - dependent)
3. Cognitive style (analytical - intuitive)

We can analyze the individual behavior of the traders regarding this, as shown in **Table 17**.

**Table 17**. Agent's earnings regarding his or her psychological behavior.

| **Trader** | **Experts** | | **Analytic** | | **Calm** | | **Dependent** | |
|---|---|---|---|---|---|---|---|---|
| **Trader** | Yes | No | Yes | No | Yes | No | Yes | No |
| **1** | -15,12 | | -15,12 | | -15,12 | | -15,13 | |
| **2** | 81,53 | | 81,53 | | 81,53 | | | 81,53 |
| **3** | -6,78 | | -6,78 | | | -6,78 | -6,78 | |
| **4** | 74,62 | | 74,62 | | | 74,62 | | 74,63 |
| **5** | 44,09 | | | 44,09 | 44,09 | | | 44,09 |
| **6** | 76,03 | | | 76,03 | 76,03 | | 76,03 | |
| **7** | 20,78 | | | 20,78 | | 20,78 | | 20,78 |
| **8** | 10,59 | | | 10,59 | | 10,59 | 10,59 | |
| **9** | | -26,56 | -26,56 | | -26,56 | | -26,56 | |
| **10** | | 74,65 | 74,65 | | 74,65 | | | 74,66 |
| **11** | | -88,71 | | -88,71 | -88,71 | | -88,72 | |
| **12** | | 103,93 | 103,93 | | | 103,93 | 103,94 | |
| | **35,71** | **7,91** | **40,89** | **8,96** | **20,84** | **29,02** | **7,63** | **42,24** |

We can analyze this table, focusing on each of these independent features. In that sense, expert traders and analytical traders perform better than other traders. Also, the difference between calm and anxious (calm=”no”) traders is small when compared with the differences of other features.
An interesting point is the fact that traders that have less social dependency (independent traders 2, 4, 5, 7 and 10) always earn. This only happens in this feature, not in the Expert/Non-expert feature or Analytic/Non-analytic feature, since here there are cases in which the agent does not earn. An example is agent 9, who is an analytical (and calm and dependent) trader that does not earn in every scenario. However, as we can see next, we cannot conclude that being independent is a key factor for winning. Therefore, more research is needed in that sense. To analyze this in detail, we compare the earnings of different traders for each of the rounds terms of his or her knowledge of the value of the asset and his or her psychological features. We use the Wilcox test (see Table 18). To see the complete analysis, Annex 1 can be reviewed. The hypothesis used in each test is that the median of both distributions are equal, and the alternative is that they are different, using a 10% level.

Table 18. Summary of the test performed and conclusions obtained for the different trader populations.

| Test | | p-value | Conclusion |
|---|---|---|---|
| Non-informed | Expert vs. non-expert | 0.05808 | Reject |
| | Analytic vs. non-analytic | 0.0002999 | Reject |
| | Calm vs. no calm | 0.7615 | Do not reject |
| | Dependent vs. non-dependent | 0.07441 | Reject |
| Informed | Expert vs. non-expert | 0.1829 | Do not reject |
| | Analytic vs. non-analytic | 0.3256 | Do not reject |
| | Calm vs. not calm | 0.8874 | Do not reject |
| | Dependent vs. non-dependent | 0.1904 | Do not reject |

It is not a surprise that, in our model, important psychological factors for a trader are *analytic*, *dependent*, and *expert*. For an informed agent, the psychological factor seems to be non-determinant. However, for a non-informed agent, *analytic* becomes the main factor for success. Also, it is quite interesting that our model behaves as expected for the informed traders, and that no psychological feature seems to affect the overall trading behavior very much. This is because the trader knows the real value of the asset when he or she is informed.
Regarding the limitations of the model, note that we are not using randomness; this implies that the scenarios always perform identically. We choose not to use randomness in order to simplify the executions, focusing our efforts on the Conceptual Validation and Verification of the model, see (Sargent, 2007).
Regarding the validation process, it is true that substantial complexity is involved in ensuring that a model is valid simply because it behaves the same as the system it tries to represent. This problem is more complex and difficult to solve with social simulation models (see (Küppers & Lenhard, 2005) for a complete review of this problem). However, here we are presenting a methodology that works because it simplifies the Conceptual Validation and the automatic implementation of the model (from this conceptual model).
With this project, we have used a brief but significant number of well-selected features to simulate the behavior of a trader, and we have obtained results similar to those of a real trader. These features can be used in other MABS models to represent a detailed trader’s behavior. From our experiment, we agree with the Veiga and Vorsatz (Veiga & Vorsatz, 2008) experiment, in the sense that we reject the assumption of the efficient market (Hayek, 1945), which states that the price of an asset summarizes all information available to market participants or, in other words, that markets are immune to price manipulation. We see in our experiment that this is not generally true, and that in certain circumstances the price can be manipulated.
Finally, one of the main objectives of the paper is to show that the proposed methodology, which is based on the use of SDL diagrams, is suitable for describing social models, specifically the behavior of traders. Using SDLPS as a tool for its implementation simplifies the verification process and accelerates the final result. Its ability to work at a high level, requiring only the definition of a model flow diagram for simulation, simplifies the validation process for users with limited programming skills, and it provides a mechanism for faster creation of lean simulation models and prototypes.

## 9. Statements & Declarations

Funding. The authors declare that no funds, grants, or other support were received during the preparation of this manuscript.
Competing Interests. The authors have no relevant financial or non-financial interests to disclose.

## 10. Acknowledgements

Many thanks to Aaron Montero, who defined a preliminary model; and to Manuel Montero, without whose expertise and guidance in defining the psychology of the agents, this work could not have been done.

## 11. Compliance with Ethical Standards

Conflict of Interest: The authors declare that they have no conflict of interest.

## 12. References

Allport, G. W. (1963). *Pattern and Growth in Personality.* Harcourt College Publishers.

Ballesteros, R. F. (1984). *Psicodiagnóstico.* UNED.

Ballesteros, S. (2000). *Psicología General: Un enfoque cognitivo* (4 ed.). Universitas.

Ballesteros, S., & García, B. (2005). *Procesos psicológicos básicos.* Universitas.

Brandstätter, H. (1993). Should economic psychology care about personality structure? *Journal of Economic Psychology, 14*, 473-494.

Cárpatos, J. L. (2006). *Leones contra gacelas. Manual completo del especulador.* (4 ed.). Millennium Capital, S.L.

Cattell, R. B. (1966). The Scree Test for the Number of Factors. *Multivariate Behavioral Research, 1*(2), 245-276.

Cattell, R. B., Eber, H. W., & Tatsuoka, M. M. (1980). *Handbook for the Sixteen Personality Factor Questionnaire (16 PF).* Champaign IL: Institute for Personality and Ability Testing.

Davidsson, P. (2001). Multi Agent Based Simulation: Beyond Social Simulation. (S. Moss, & P. Davidsson, Eds.) *Lecture Notes in Computer Science, 1979*, 97-107. doi:10.1007/3-540-44561-7_7

Doldi, L. (2001). *SDL illustrated - visually design executable models.* TRANSMETH SUD OUEST.

Doldi, L. (2003). *Validation of Communications Systems with SDL: The Art of SDL Simulation and Reachability Analysis.* John Wiley & Sons, Inc.

Dow, C. H. (1844). *Wall Street Journal*.

Eysenck, H. J. (1991). *EPQ-R.* TEA.

Fonseca i Casas, P. (2008). SDL distributed simulator. *Winter Simulation Conference 2008.* Miami: INFORMS.

Fonseca i Casas, P., & Casanovas, J. (2011). Towards a SDL-DEVS Simulator. *The Third International Conference on Advances in System Simulation.* Barcelona.

Fonseca, P. (2012). Enhancing SDLPS with Co-simulation. In C. Laroque, J. Himmelspach, R. Pasupathy, O. Rose, & A.M. Uhrmacher (Ed.), *Proceedings of the 2012 Winter Simulation Conference.* Berlin: IEEE.

Gardner, H. (2006). *Multiple intelligences: new horizons.* Basic Books.

Hayek, F. (1945, September). The Use of Knowledge in Society. *The American Economic Review, 4*(XXXV), 519-30.

Hoffmann, A. O., Jager, W., & Von Eije, J. H. (2007). Social Simulation of Stock Markets: Taking It to the Next Level. *Journal of Artificial Societies and Social Simulation, 10*(2), 7. Retrieved October 01, 2011, from http://jasss.soc.surrey.ac.uk/10/2/7.html

IBM. (2009). *TELELOGIC*. Retrieved 03 31, 2009, from http://www.telelogic.com/

ITU-T. (2012). *Specification and Description Language (SDL).* Retrieved November 2012, from Series Z: Languages and general software aspects for telecommunication systems.: http://www.itu.int/ITU-T/studygroups/com17/languages/index.html

Kahneman, D., & Tversky, A. (1972). Subjective probability: A judgment of representativeness. *Cognitive Psychology, 3*, 430-454.

Küppers, G., & Lenhard, J. (2005). Validation of Simulation: Patterns in the Social and Natural Sciences. *Journal of Artificial Societies and Social Simulation, 8*(4).

Langer, M., Oster, D., Speith, T., Hermanns, H., Kästner, L., Schmidt, E., . . . Baum, K. (2021). What do we want from Explainable Artificial Intelligence (XAI)? – A stakeholder perspective on XAI and a conceptual model guiding interdisciplinary XAI research. *Artificial Intelligence, 296,*. doi:10.1016/j.artint.2021.103473

Langley, P. W., Bradshaw, G., Simon, H. A., & Zytkow, J. M. (1987). *Scientific Discovery: Computational Explorations of the Creative Processes.* The MIT Press.

LeBaron, B., Arthur, W. B., & Palmer, R. (1999). Time series properties of an artificial stock market. *Journal of Economic Dynamics and Control, 23*, 1487-1516.

Lo, A. W., Repin, D. V., & Steenbarge, B. N. (2005). Fear and Greed in Financial Markets: A Clinical Study of Day-Traders. *Papers and Proceedings of the One Hundred Seventeenth Annual Meeting of the American Economic Association. 95*, pp. 352-359. Philadelphia, PA: American Economic Association. Retrieved from http://www.jstor.org/stable/4132846

Mehlenbacher, A. (2009). Multiagent System Simulations of Treasury Auctions. *Computational Economics*, 67-117. doi:10.1007/s10614-008-9165-z

Morales, J. F., Moya, M. C., Gaviria, E., & Cuadrado, I. (2007). *Psicología Social* (3 ed.). McGraw-Hill.

Moreno Bermúdez, J. (2000). *Psicología de la Personalidad.* UNED.

Nelson, S. (2007). *The ABC of Stock Speculation.* Cosimo Classics.

OMG SysML. (2010, June). *OMG SysML.* Retrieved December 2010, from OMG: http://www.omg.org/spec/SysML/1.2/

Palomo González, A. M. (1989). Laurence Kohlberg: Teoría y práctica del desarrollo moral en la escuela. *Revista interuniversitaria de formación del profesorado, 4*, 79-90.

Peterson, C., Maier, S. F., & Seligman, M. E. (1995). *Learned Helplessness: A Theory for the Age of Personal Control.* New York: Oxford University Press.

Petri, C. A. (1962). *Kommunikation mit Automaten.* Bonn: University of Bonn.

Pitarque, A., & Algaradel, S. (1991). El conexionismo como marco de simulación: Aplicación a una tarea de facilitación semántica. *Cognitiva, 3*(2), 165-186.

PragmaDev SARL. (2012). Retrieved from PragmaDev - Code generation: http://www.pragmadev.com/product/codeGeneration.html

Precher, R. R., & Frost, A. J. (1989). *El principio de la onda elliot.* (Gesmovasa, Ed.)

Rabin, M. (2002). A perspective on psychology and economics. *European Economic Review, 46*, 657 - 685.

Rayner, N., Phelps, S., & Constantinou, N. (2014). Learning is neither sufficient nor necessary: An agent-based model of long memory in financial markets. *AI Communications, 27*(4), 437-452. doi:10.3233/AIC-140608

Rogers, C., & Kramer, P. D. (1995). *On Becoming a Person: A Therapist's View of Psychotherapy.* Mariner Books.

Rotter, J. B. (1990). Internal versus external control of reinforcement: A case history of a variable. *American Psychologist, 45*, 489–93. doi:10.1037/0003-066X.45.4.489

Russell, S., & Norvig, P. (2011). *Artificial Intelligence: A Modern Approach* (3rd ed.). Prentice Hall.

Sandrila Ltd. (2009). *Sandrila SDL*. Retrieved 08 19, 2012, from Sandrila: http://www.sandrila.co.uk/visio-sdl/index.php

Sargent, R. G. (2007). Verification and Validation of simulation models. In S. G. Henderson, B. Biller, M.-H. Hsieh, J. Shortle, J. D. Tew, & R. R. Barton (Ed.), *Proceedings of the 2007 Winter Simulation Conference.* IEEE.

Silva Suárez, M. (1985). *Las Redes de Petri: en la Automática y la Informática.* Madrid: Editorial AC, D.L.

Tajfel, H. (1984). *Grupos humanos y Categorías sociales.* Herder.

Takahashi, H., & Terano, T. (2003). Agent-Based Approach to Investors' Behavior and Asset Price Fluctuation in Financial Markets. *Journal of Artificial Societies and Social Simulation, 6*(3). Retrieved October 01, 2011, from http://jasss.soc.surrey.ac.uk/6/3/3.html

Veiga, H., & Vorsatz, M. (2008). Price manipulation in an experimental asset market. *European Economic Review, 53*(3), 327-342.

Wan, H. A., Hunter, A., & Dunne, P. (2002). Autonomous Agent Models of Stock Markets. *Artificial Intelligence Review, 128*(17), 87.

Wang, D., Quek, C., & Ng, G. S. (2016). Bank failure prediction using an accurate and interpretable neural fuzzy inference system. *AI Communications, 29*(4), 477-495. doi:10.3233/AIC-160702

Withanawasama, R., Whighama, P., & Crack, T. (2013). Characterising trader manipulation in a limit-order driven market. *Mathematics and Computers in Simulation, 93*, 43–52. doi:10.1016/j.matcom.2012.09.012

Yerkes, R. M., & Dodson, J. D. (1908). The relation of strength of stimulus to rapidity of habit-formation. *Journal of Comparative Neurology and Psychology, 18*, 459–482.

Zeigler, b., Praehofer, h., & Kim, d. (2000). *Theory of Modeling and Simulation.* Academic Press.

## 13. Annexes

### 13.1. Agents SDL specification

#### 13.1.1. Market block

This block consists of a single process, *P_Market* which has the same channels as seen before for the block *B_Market* which are a bridge between the inner workings of the block and other system agents (see Fig. 8).

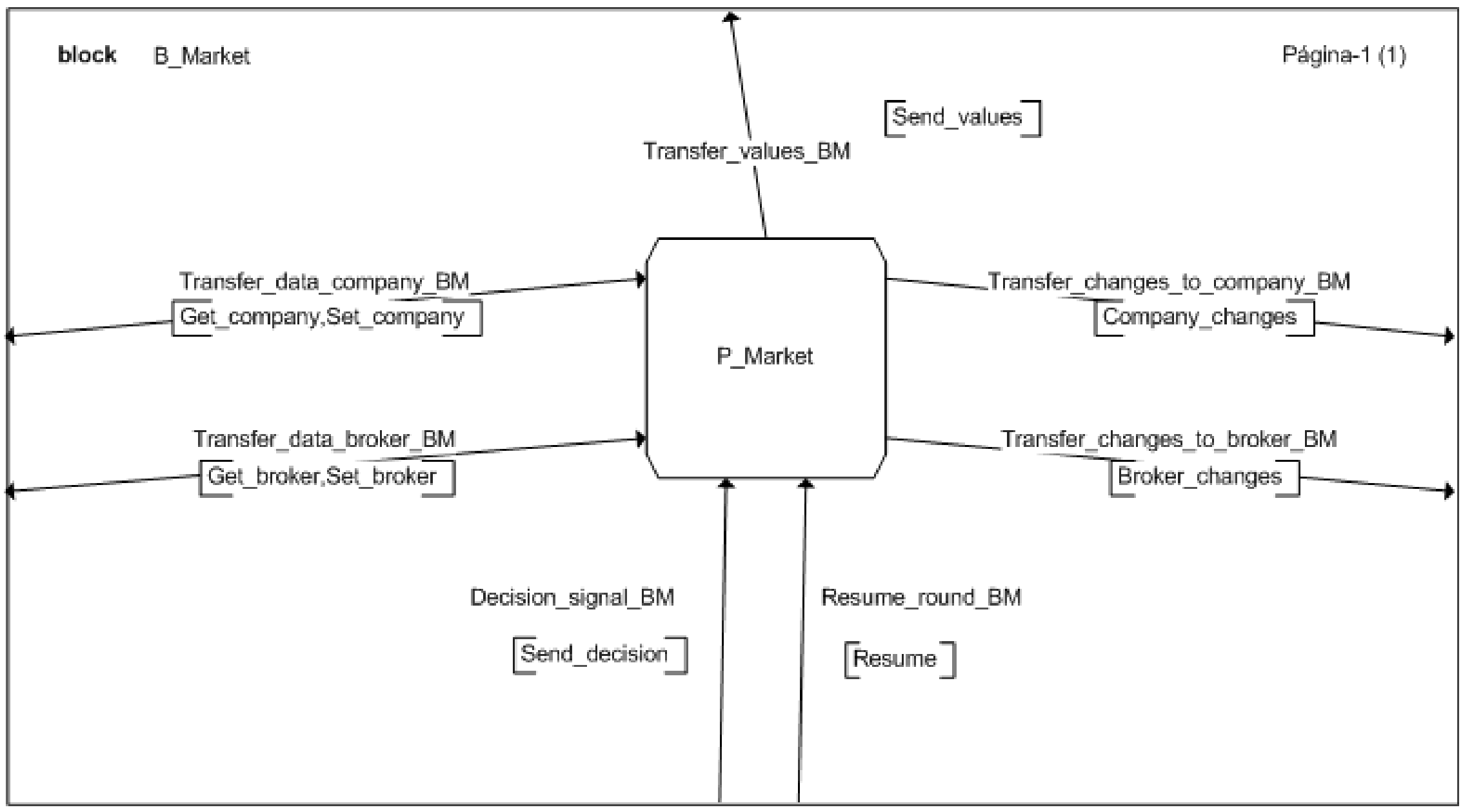

Fig. 8. B_Market block.

#### 13.1.2. Company block

This block, like block *B_Market* only made up of a process *P_Company* which receives the full channels linked to the block *B_Company* (see Fig. 9).

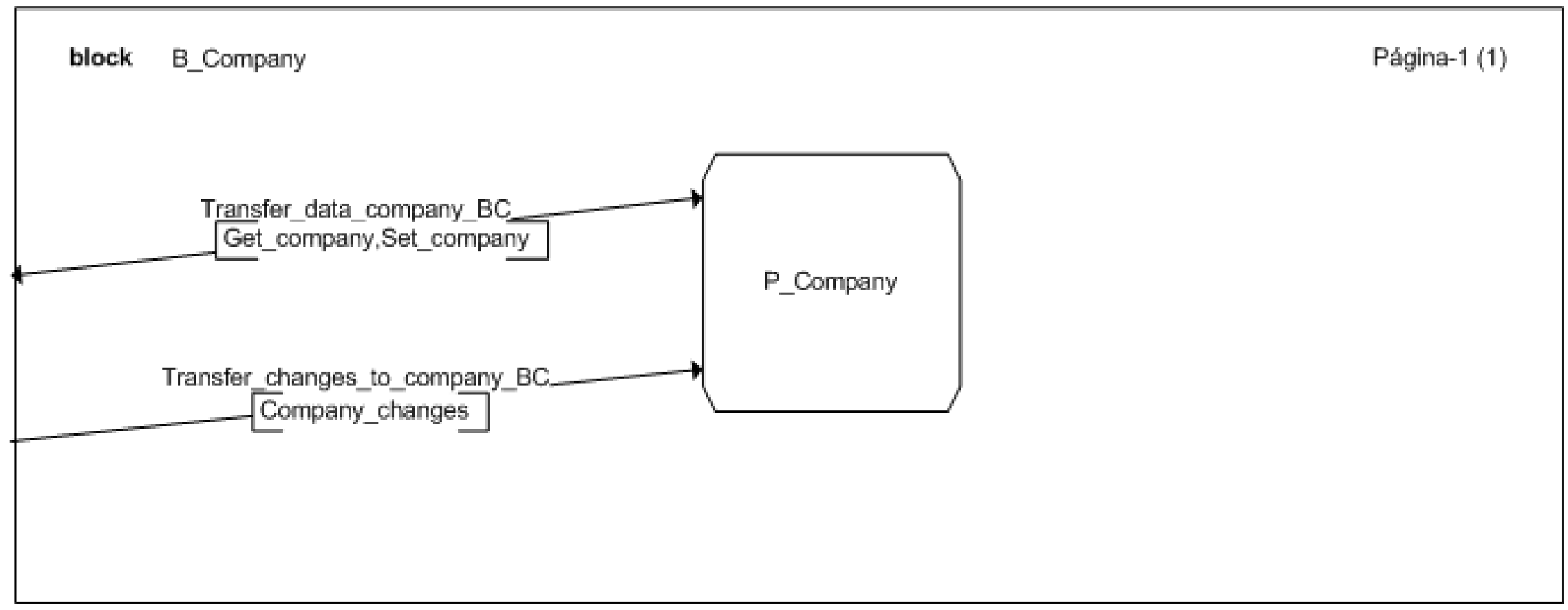

Fig. 9. B_Company block.

#### 13.1.3. Broker block

This block, like the rest of the blocks previously seen, is composed of a single process *P_Broker* which consists of the same channel that owns the block *B_Broker* to communicate with the system (see Fig. 10).

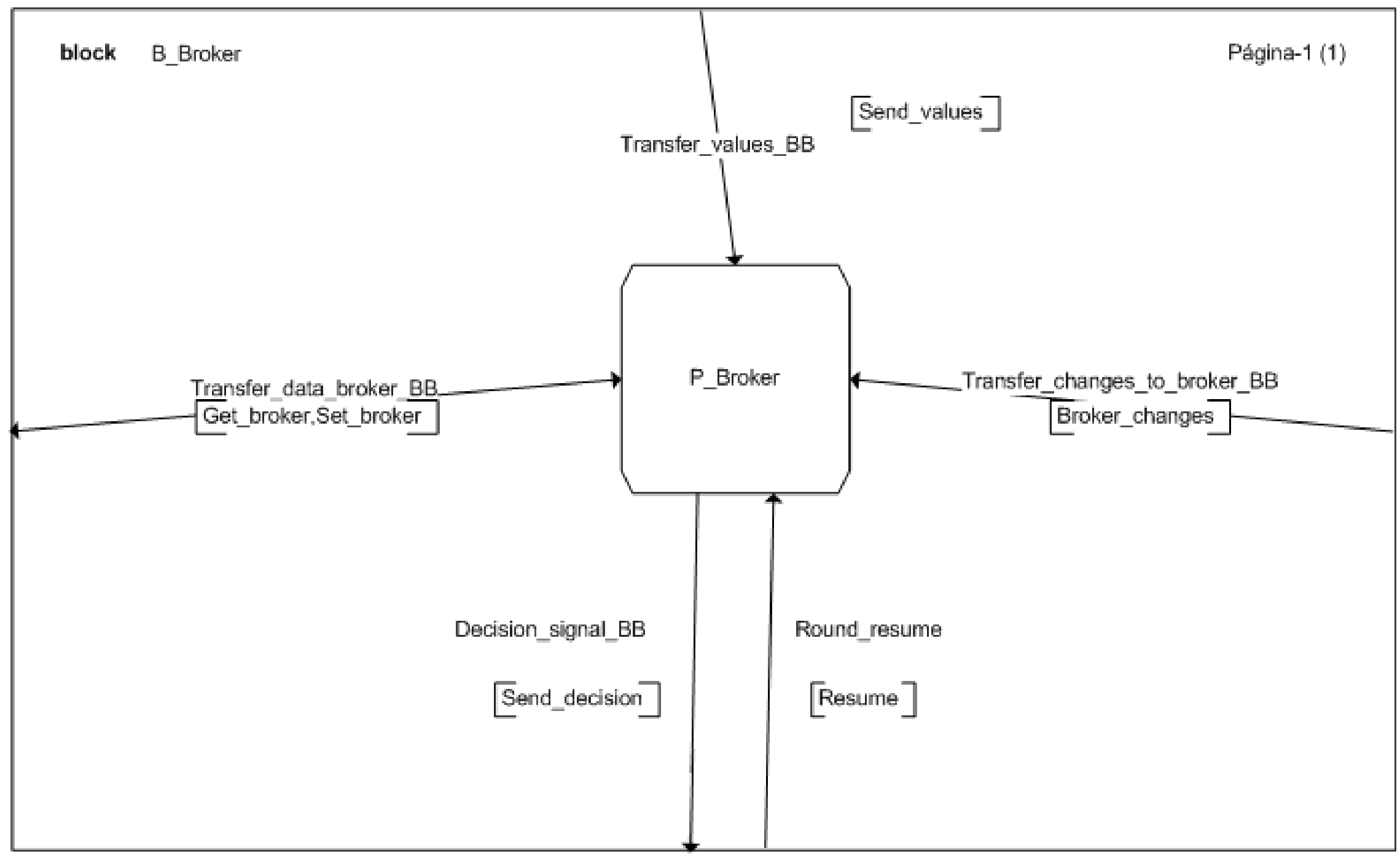

Fig. 10. B_Broker block.

#### 13.1.4. Market process

The process *P_Market* is the core of our system, responsible for communicating to other processes, analyzing and processing the information that arrives from the other agents and transmitting the result of this processing. That is why this process is where lies our system clock, counting the time and transmitting that information to other processes. The clock is built using a procedure, *Pr_Market_time* after which emits a signal *Flux_time* which will be collected as an input signal through the same process *P_Market*. When the signal is received the process begins anew. This creates a cyclical process that will allow us to control the timer.

Once created the counter, called *timer*, consider whether or not we are at the beginning of a new turn. If that is so, the simulation did not come to an end and we initialize the relevant variables, but if the simulation should come to an end, we turn to the state well, *Terminate*. If we have not begun a new turn, the process determines which brokers are best sellers and buyers in the market, from the data we have from them. And once established this information, it selects the best buyer and seller for each broker for the market value. This information, along with the timer and if the broker has been informed in this round, is passed to the broker using the sign *Get_broker*. With this we seek to obtain information from a particular broker. In turn, we send the signal *Get_company* with only the system timer, pursuing a similar goal to the previous signal, only in the case of the company.

From this point the process *P_Market* shall receive external signals to be processed.

In the case of the signals *Set_broker* and *Resume* the process *P_Market* only update the values it has. First it updates the values of a particular broker with the first signal. With the second the final information received from a particular broker, credit, equity, and bottom line.

For signal *Set_company* the process *P_Market* updates the values of the company that are communicating with it, but in turn transmits these values to the various brokers that are in the system via the signal *Send_values*.

Finally, it remains to discuss the case of signal *Send_decision* which implies a higher level of processing of the data received. This signal means in the case that the decision of the broker is buying or selling, to determine what the rest of the brokers is currently the best buyer or seller and assess whether it is possible to close a deal with it. If so, close the deal and pass the updated values to the brokers and company involved through the signals *Broker_changes* and *Company_changes* (actually *Broker_X_changes* and *Broker_Y_changes* where X and Y are identifiers the broker). But if the deal cannot close, we inform the broker that sent us his decision, which it has not, sought the fruits obtained by *Broker_changes* signal. So the broker can assimilate this information so that the agent can succeed next time.

The state *Terminate* is a state well for one simple reason, once we are in it, for any signal we receive, (* means any signal) will not perform any operation, only staying in the same state. So all new signals is ignored until the simulation model knows she has completed and finished (see Fig. 11).

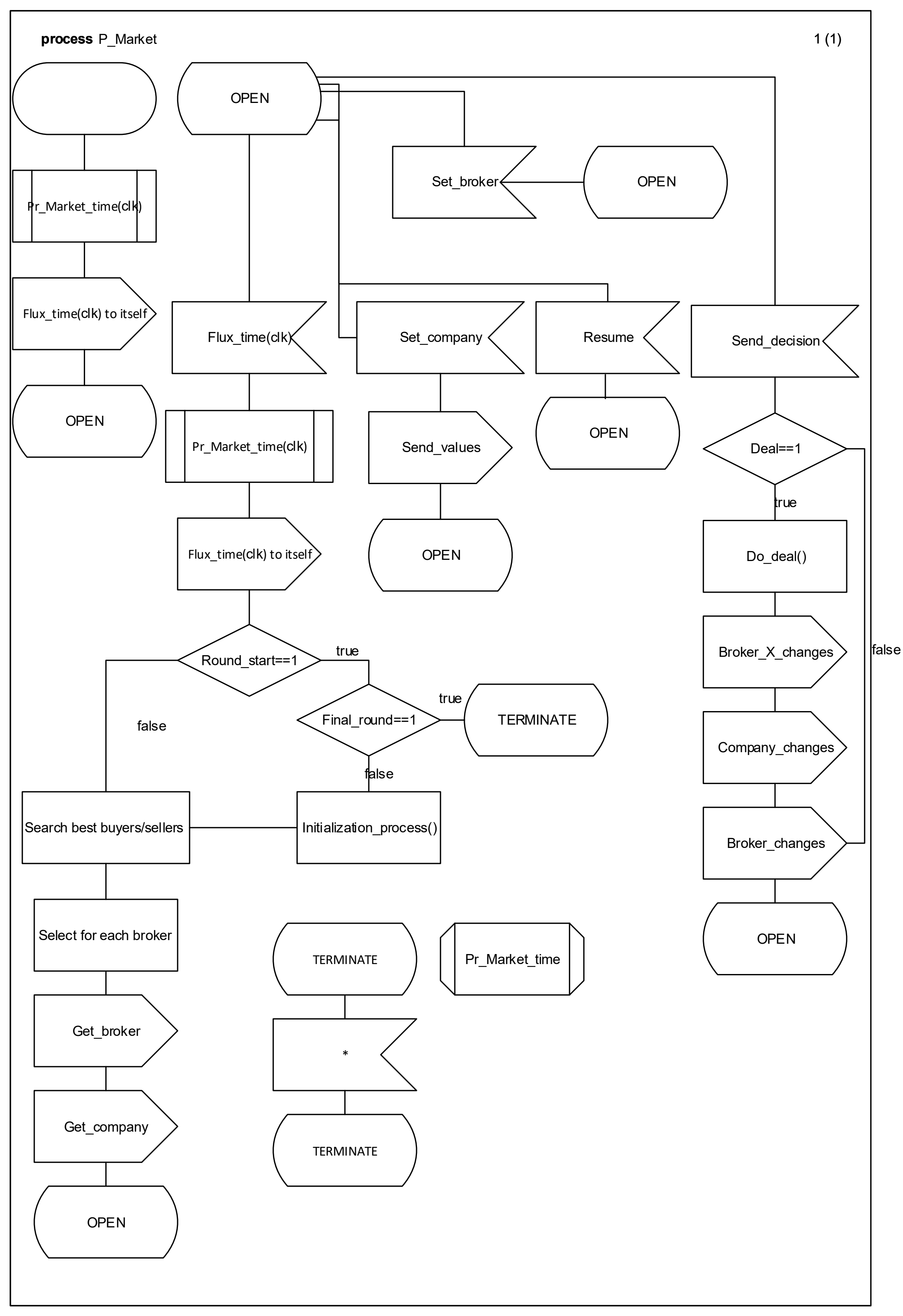

Fig. 11. Process *P_Market*. The procedure *Pr_Market_time* is only a process that assigns a unit of time to a variable named market_clk, representing the simulation clock.

#### 13.1.5. Broker process

The process *P_Broker* receives three external signals, *Get_broker*, *Send_values* and *Broker_changes*.
First signal checks if you have begun a new round or not. If you start a new round, the agent sees whether or not a previous round exists, and if so whether it has reached the final round or not. If you have reached the end of the round, as happens in other processes, we send our process to the state well, *Terminate*. However, if the simulation is complete but there is a previous round, we will need to calculate the final results of the round, essentially benefits, and transmit these, with credit and number of shares the broker has to market. In any case, whenever you start a new round, we should proceed to set the values of brokers, consisting mainly in initializing and calculating their values from other information received from the market, if a broker is informed or not, and the other values that the broker just initialize. In any case we send the *Set_broker* signal with the information belonging to the broker to the market.
With the arrival of the signal *Send_values* will begin the more interesting and complex process of our simulation model, the one who set up the decisions taken by the broker from the receiving market information. This decision making process is that we really want to simulate in our model, for that is why it has been the process to which we has devoted more attention, trying to describe as clearly and concisely as possible.
The first thing we must resolve, once we receive the signal *Send_values,* is whether the broker, that we are going to simulate, has the human, robot or manipulator agent behavior.
In the case that the broker is the robot, the operations flow is simple. Only we will follow the strategy set by the article of Veiga and Vorsatz (Veiga & Vorsatz, 2008).
In the case that the agent represents a human behavior, we will first see if the broker is ready to perform an operation. If it is, we will calculate the next delay (ie, how long it takes to make a new operation) and proceed to build a decision from the information provided to us in the market.
To establish such a decision, first we will see if our broker is experienced or not, and if so, to consider whether there are signs of buying or selling in the market. The next thing is find out the trend that follows the company for which we will decide whether to buy, sell or issue any further action. And from this tendency, plus the information received from the market, to establish a prediction about the future of their actions.
However, not always have the information needed to predict the future value of a company, especially when we are at the beginning of the simulation. On that occasion, we will make a decision based on what the broker believes that will happen, hunch, or secure information (in the case that the broker has been informed of the real value of the market action).
In any case, the broker will make a decision, and if this is a purchase or sale, should bear in mind the demands of supply and demand in the market already issued, in order to make an offer to buy or sell securities consequent with the promulgated values.
Finally established which will be the brokers decision, purchase or sale, it is needed to establish which will be the value of the offer and the volume of shares involved. Next it is needed to verify what we can deal with this decision and correct it, if is necessary. Once we can deal with the decision, we transmit the signal to the market through *Send_decision*, waiting to find a buyer or seller, if our desire is to participate actively in the market.
Lastly, with the arrival of the signal *Broker_changes*, we update the values of the broker, issued from the market. These signals are being processed only if the broker is in *Trade* state because, being in the state well, *Terminate*, we will ignore it, as we have done in other processes.
This process is represented on Fig. 12.

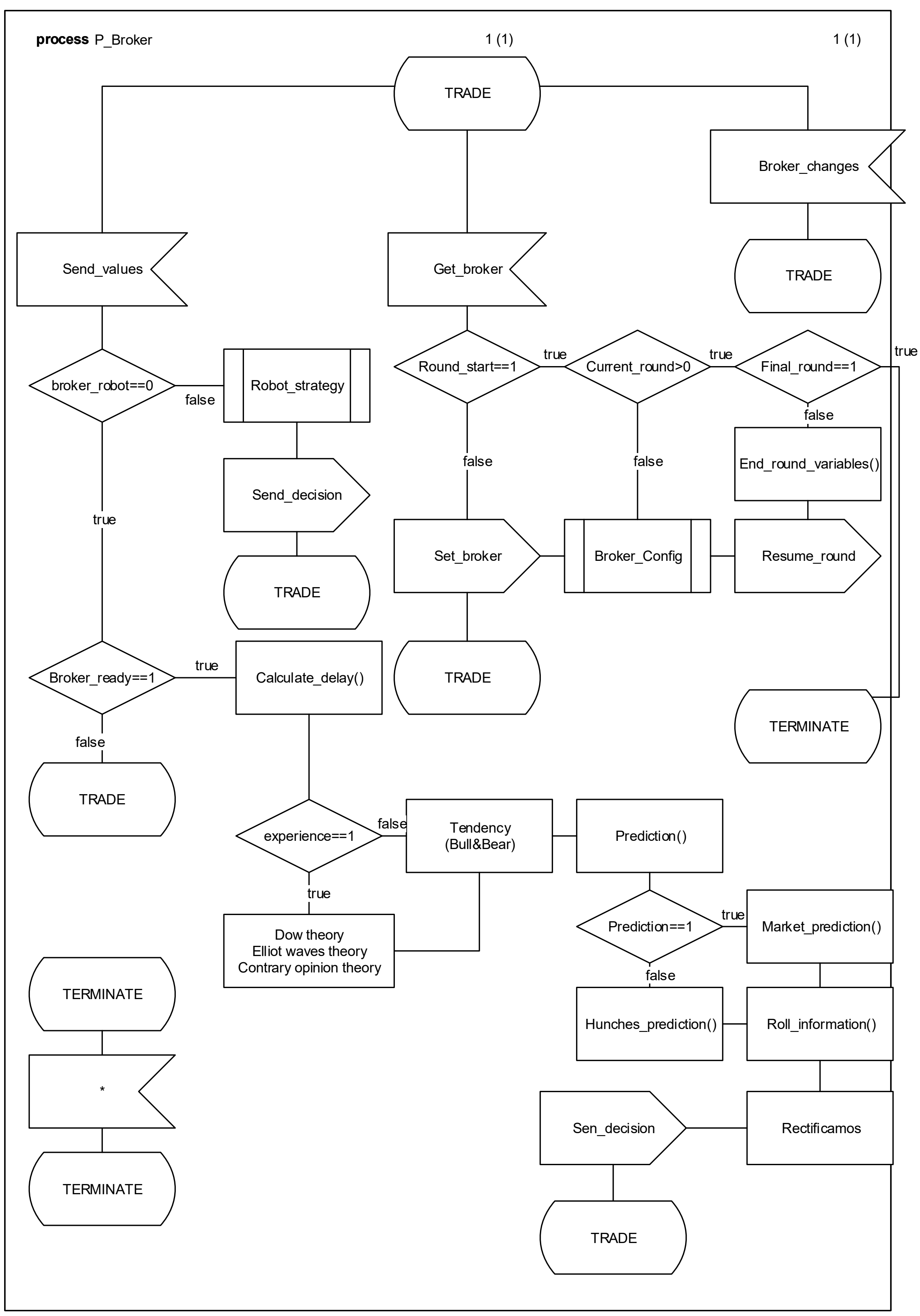

Fig. 12. The figure shows the process defining the broker behavior for TRADE state.

## 13.2. Preliminary analysis of the relation of the trader traits and the benefits

During the validation process as a guidance, we analyse the relation of the different factors regarding the benefits.

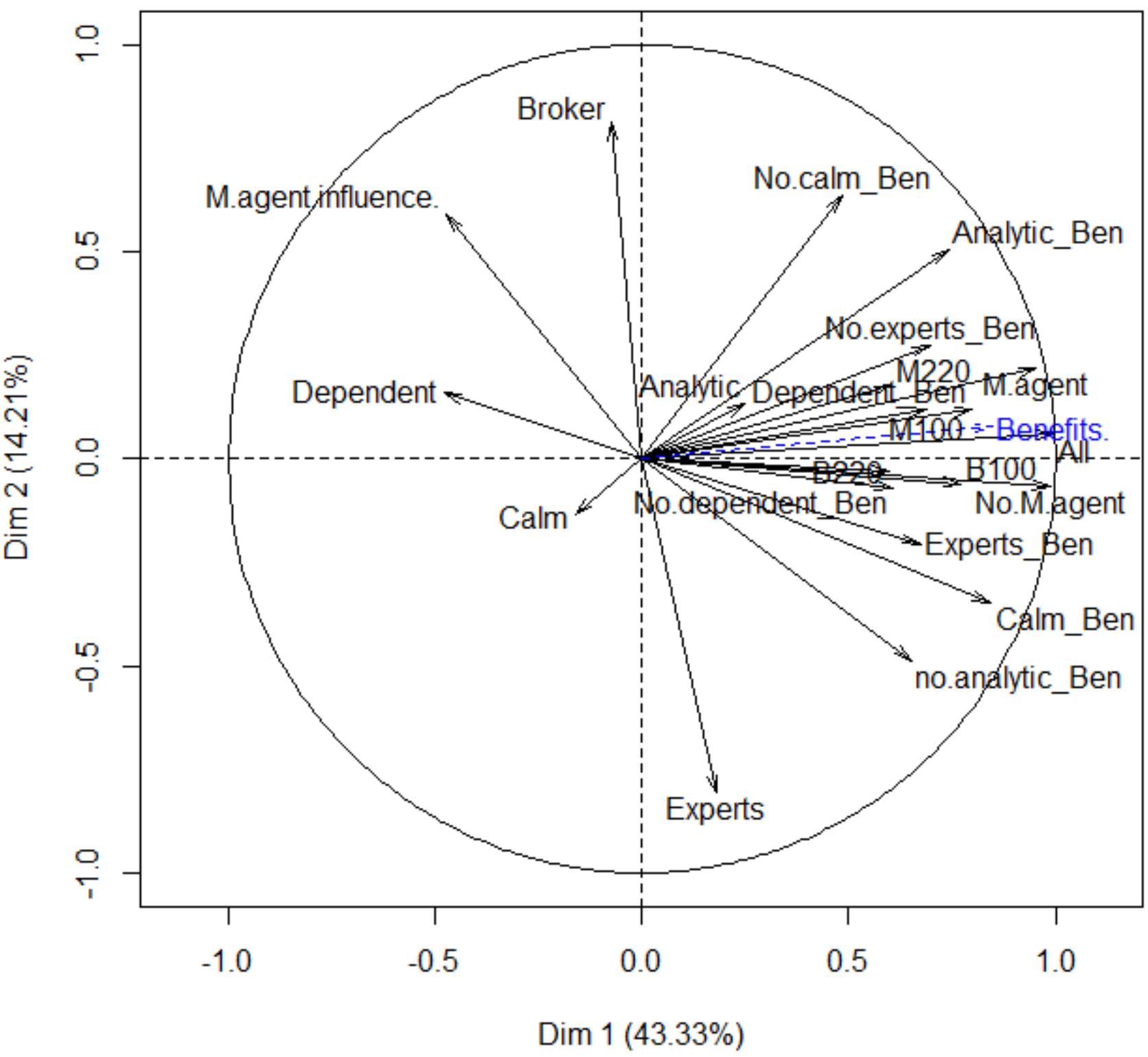

Analytic and no-Dependent are the main traits that we desire in order to improve the overall benefits, becoming no-Dependent one of the main trait to be considered.

Table 19. Effect of the “dependent” psychological feature for no informed traders. The two distributions have not so different medians (true location shift is not equal to 0).

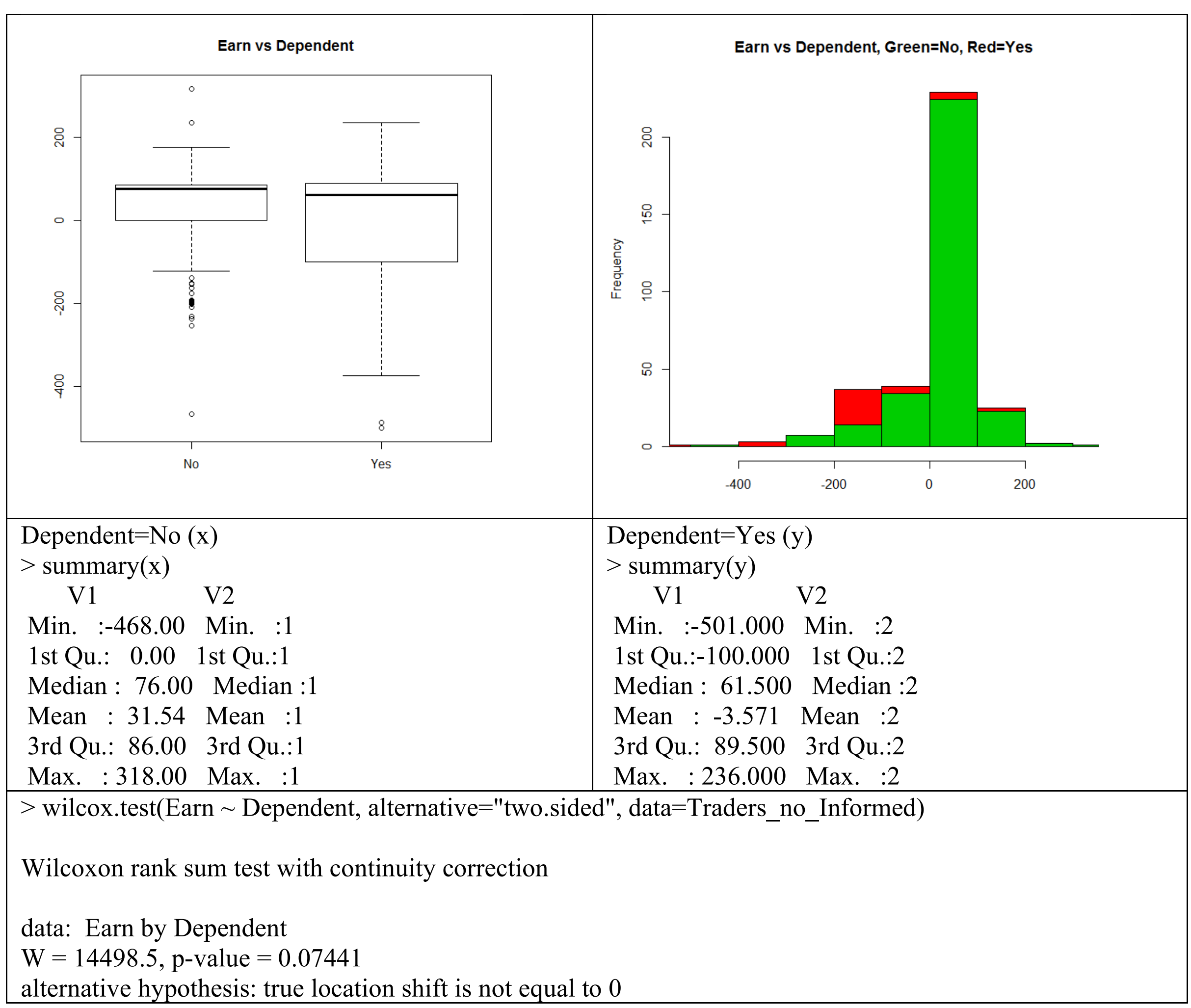

| Dependent=No (x) | Dependent=Yes (y) |
|---|---|
| > summary(x) | > summary(y) |

```
      V1              V2
 Min.   :-468.00   Min.   :1
 1st Qu.:   0.00   1st Qu.:1
 Median :  76.00   Median :1
 Mean   :  31.54   Mean   :1
 3rd Qu.:  86.00   3rd Qu.:1
 Max.   : 318.00   Max.   :1
```

```
      V1               V2
 Min.   :-501.000   Min.   :2
 1st Qu.:-100.000   1st Qu.:2
 Median :  61.500   Median :2
 Mean   :  -3.571   Mean   :2
 3rd Qu.:  89.500   3rd Qu.:2
 Max.   : 236.000   Max.   :2
```

```
> wilcox.test(Earn ~ Dependent, alternative="two.sided", data=Traders_no_Informed)

Wilcoxon rank sum test with continuity correction

data:  Earn by Dependent
W = 14498.5, p-value = 0.07441
alternative hypothesis: true location shift is not equal to 0
```

Table 20. Effect of the “expert” psychological feature for informed traders. The two distributions have not so different medians (true location shift is not equal to 0).

| Earn vs Expert | Earn vs Expert, Green=No, Red=Yes |
|---|---|

```
Expert=No (x)
> summary(x)
       V1              V2
 Min.   :-146.00   Min.   :1
 1st Qu.:   0.00   1st Qu.:1
 Median :  32.00   Median :1
 Mean   :  29.79   Mean   :1
 3rd Qu.:  82.00   3rd Qu.:1
 Max.   : 135.00   Max.   :1
```

```
Expert=Yes (y)
> summary(y)
       V1              V2
 Min.   :-92.00   Min.   :2
 1st Qu.:  0.00   1st Qu.:2
 Median : 73.00   Median :2
 Mean   : 61.28   Mean   :2
 3rd Qu.:103.75   3rd Qu.:2
 Max.   :251.00   Max.   :2
```

```
> wilcox.test(Earn ~ Expert, alternative="two.sided", data=Traders_Informed)

Wilcoxon rank sum test with continuity correction

data:  Earn by Expert
W = 768.5, p-value = 0.1829
alternative hypothesis: true location shift is not equal to 0
```

Table 21. Effect of the “analytic” psychological feature for informed traders, both distributions The two distributions have not so different medians (true location shift is not equal to 0).

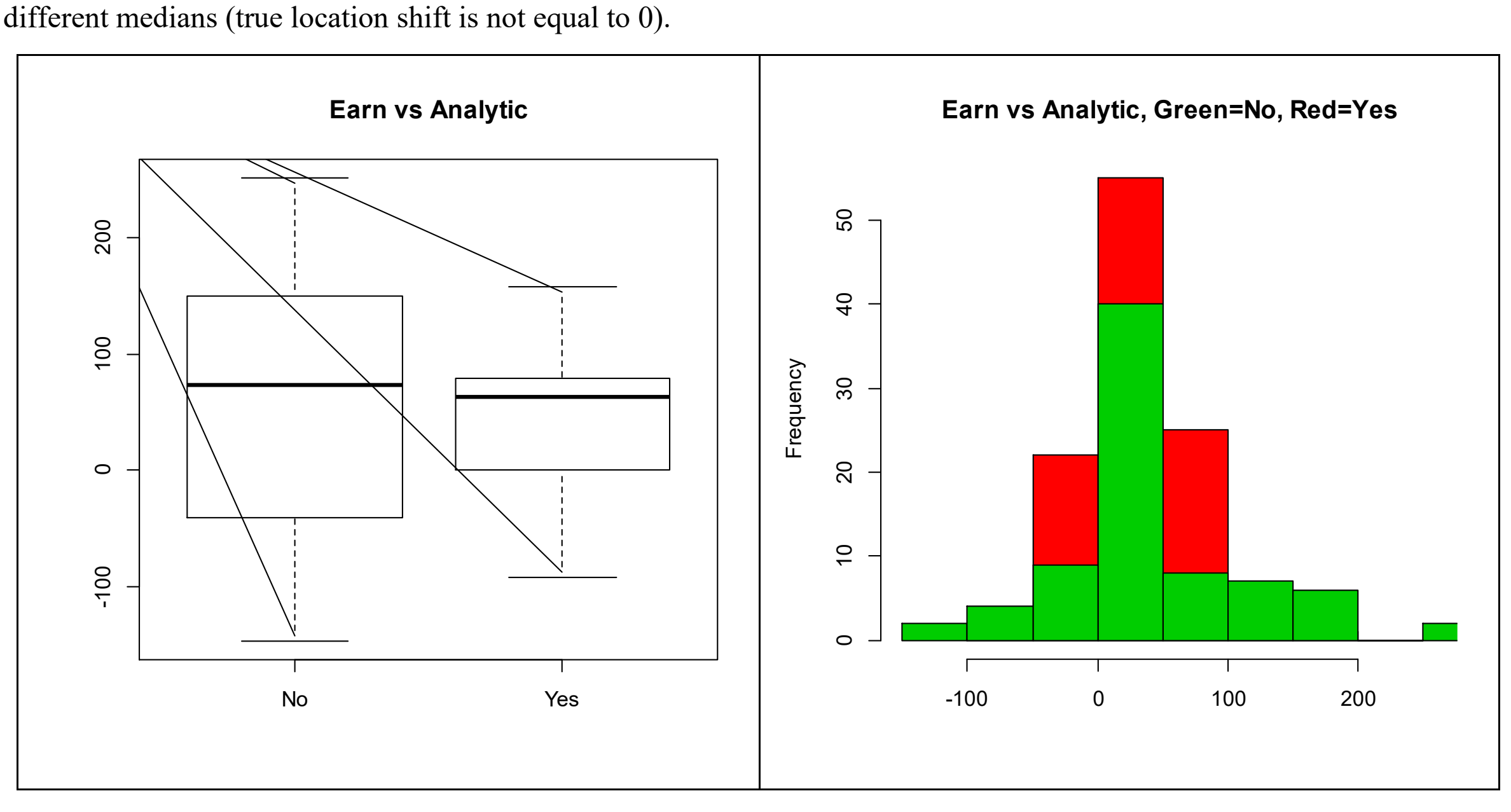

| Analytic=No (x) | Analytic=Yes (y) |
|---|---|
| > summary(dta[dta[,2]==1,])<br>V1 V2<br>Min. :-146.00 Min. :1<br>1st Qu.: -40.00 1st Qu.:1<br>Median : 74.00 Median :1<br>Mean : 58.46 Mean :1<br>3rd Qu.: 150.00 3rd Qu.:1<br>Max. : 251.00 Max. :1 | > summary(dta[dta[,2]==2,])<br>V1 V2<br>Min. :-92.00 Min. :2<br>1st Qu.: 0.00 1st Qu.:2<br>Median : 63.50 Median :2<br>Mean : 46.41 Mean :2<br>3rd Qu.: 79.00 3rd Qu.:2<br>Max. :158.00 Max. :2 |

```
> wilcox.test(Earn ~ Analytic, alternative="two.sided", data=Traders_Informed)

Wilcoxon rank sum test with continuity correction

data:  Earn by Analytic
W = 1178.5, p-value = 0.3256
alternative hypothesis: true location shift is not equal to 0
```

Table 22. Effects of the "calm" psychological feature for informed traders. The two distributions have not so different medians (true location shift is not equal to 0).

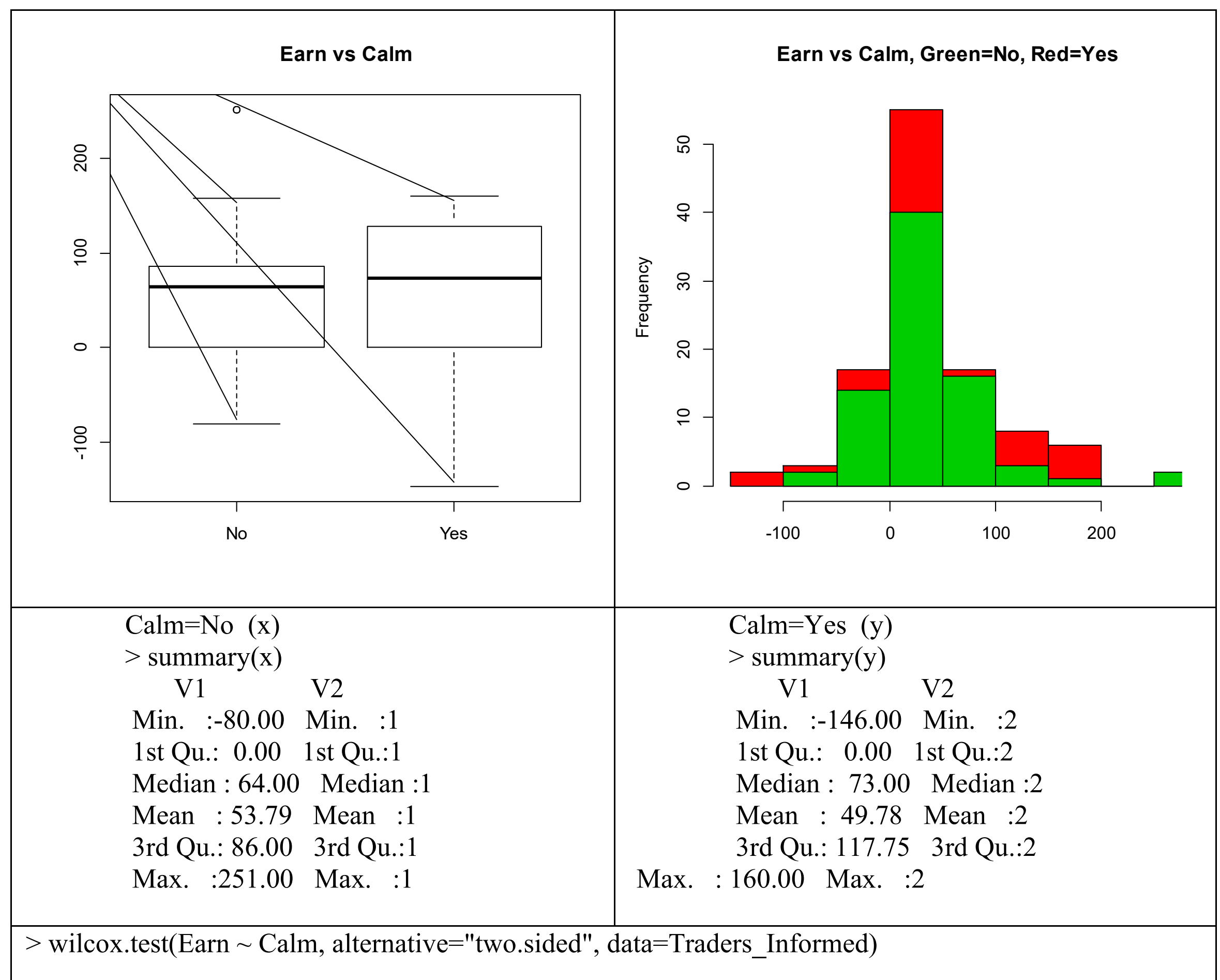

| Calm=No (x) | Calm=Yes (y) |
|---|---|
| > summary(x)<br>V1 V2<br>Min. :-80.00 Min. :1<br>1st Qu.: 0.00 1st Qu.:1<br>Median : 64.00 Median :1<br>Mean : 53.79 Mean :1<br>3rd Qu.: 86.00 3rd Qu.:1<br>Max. :251.00 Max. :1 | > summary(y)<br>V1 V2<br>Min. :-146.00 Min. :2<br>1st Qu.: 0.00 1st Qu.:2<br>Median : 73.00 Median :2<br>Mean : 49.78 Mean :2<br>3rd Qu.: 117.75 3rd Qu.:2<br>Max. : 160.00 Max. :2 |

```
> wilcox.test(Earn ~ Calm, alternative="two.sided", data=Traders_Informed)

Wilcoxon rank sum test with continuity correction

data:  Earn by Calm
W = 1034.5, p-value = 0.8874
alternative hypothesis: true location shift is not equal to 0
```

Table 23. Effect of the “dependent” psychological feature for informed traders, both distributions have different medians (true location shift is not equal to 0).

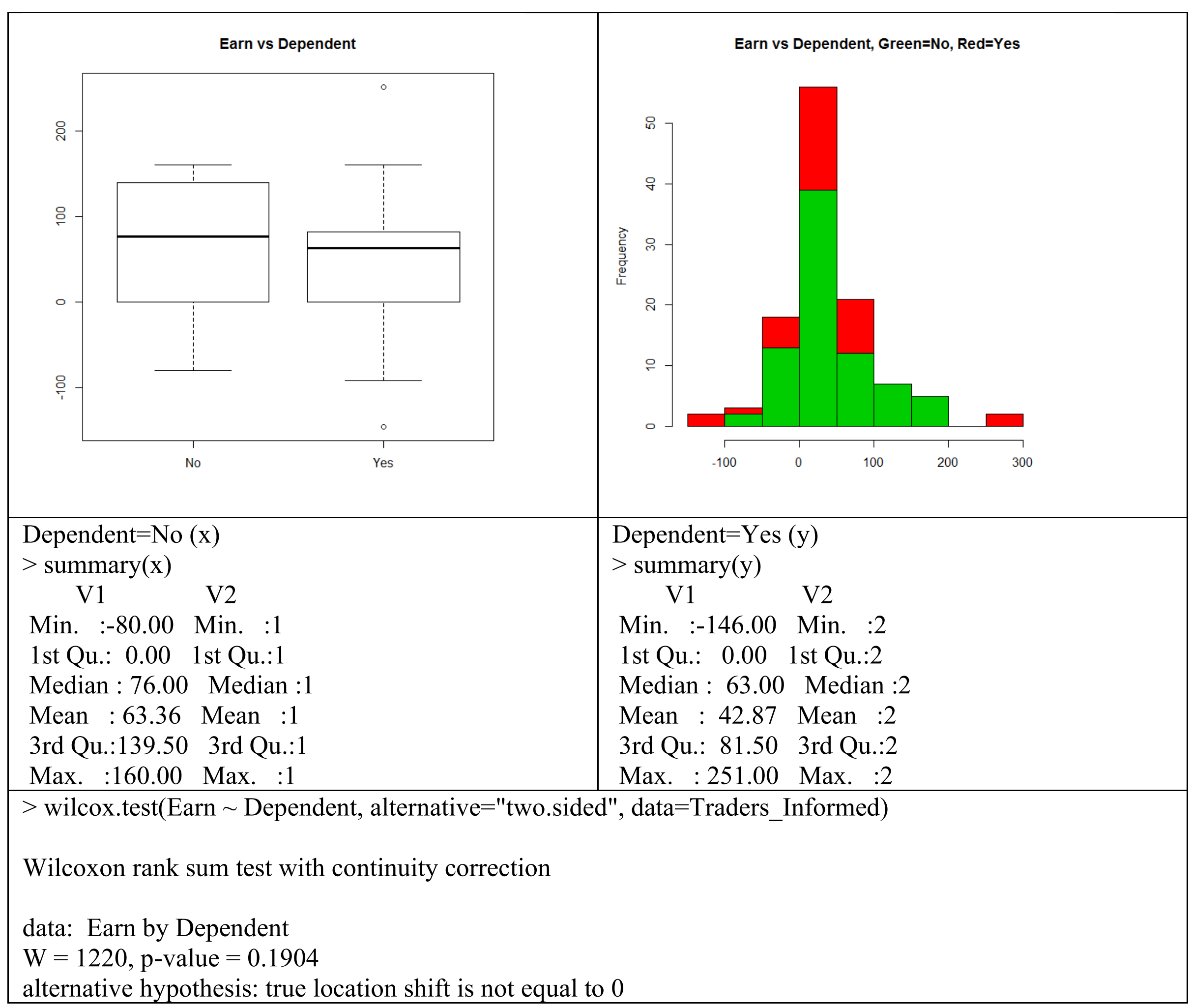

```
Dependent=No (x)
> summary(x)
       V1              V2
 Min.   :-80.00   Min.   :1
 1st Qu.:  0.00   1st Qu.:1
 Median : 76.00   Median :1
 Mean   : 63.36   Mean   :1
 3rd Qu.:139.50   3rd Qu.:1
 Max.   :160.00   Max.   :1
```

```
Dependent=Yes (y)
> summary(y)
       V1              V2
 Min.   :-146.00   Min.   :2
 1st Qu.:   0.00   1st Qu.:2
 Median :  63.00   Median :2
 Mean   :  42.87   Mean   :2
 3rd Qu.:  81.50   3rd Qu.:2
 Max.   : 251.00   Max.   :2
```

```
> wilcox.test(Earn ~ Dependent, alternative="two.sided", data=Traders_Informed)

Wilcoxon rank sum test with continuity correction

data:  Earn by Dependent
W = 1220, p-value = 0.1904
alternative hypothesis: true location shift is not equal to 0
```